\documentclass[aps,prc,floatfix,groupedaddress,showpacs,amsfonts,nofootinbib,twocolumn]{revtex4}
\usepackage{bm}
\usepackage{epsfig}
\usepackage{amsmath}
\usepackage{dcolumn}
\begin{document}

\title{The role of excited states in rp--process for sd shell nuclei}

\author{J. Grineviciute \footnote{Present address: Dept. of Physics, Western Michigan University, Kalamazoo, MI 49008, USA}}
\affiliation{National Superconducting Cyclotron Laboratory, Dept. of Physics and Astronomy, Michigan State University, East Lansing, MI 48824, USA}

\author{B. A. Brown}
\affiliation{National Superconducting Cyclotron Laboratory, Dept. of Physics and Astronomy, Michigan State University, East Lansing, MI 48824, USA}

\author{H. Schatz}
\affiliation{National Superconducting Cyclotron Laboratory, Dept. of Physics and Astronomy, Michigan State University, East Lansing, MI 48824, USA}
\affiliation{Joint Institute for Nuclear Astrophysics, Michigan State University, East Lansing, MI 48824, USA}

\begin{abstract}

We investigate the importance of proton capture on low-lying excited states that are thermally excited in hot stellar enviroments. In addition to the previously known case of $^{32}$Cl$(p,\gamma)^{33}$Ar we find several other sd-shell nuclei with large stellar enhancement factors. We discuss the uncertainty this introduces into rp-process nucleosynthesis.

\end{abstract}


\maketitle

\section{Introduction}

The rapid proton capture (rp--process) is a sequence of rapid proton capture reactions and $\beta^+$ decays passing through proton--rich nuclei. It is a dominant nucleosynthesis process in explosive hydrogen burning conditions \cite{WaW81}. Reaction rates govern energy release and final isotopic abundances. rp--process rates are crucial for X--ray burst models \cite{Sch05}. Some of the capture rates along rp--process are not determined experimentally. But many rely on theoretical input. 
Previous predictions of reaction rates have generally assumed that all interacting nuclei are in the ground state \cite{WaW81, Iliadis01, WoG94, HGW95, Vou94}.
Schatz et al. \cite{Sch05, CBB04} showed that the capture on the low-lying first-excited state of $^{32}$Cl was crucial for the $^{32}$Cl(p,$\gamma$)$^{33}$Ar reaction rate. The energies and spin in $^{33}$Ar were provided by experiment, but input for the proton and gamma decay widths come from theory. The impact of the excited states in the target nuclei that are thermally populated  in astrophysical plasma is expressed in terms of a ``stellar enhancement factor'' (SEF). The SEF for $^{32}$Cl was found to be around five for temperatures in the range 0.1 to 0.8  GK. In this paper we investigate other sd-shell nuclei where low-lying excited states may lead to large SEF.

\section{Reaction rates}

In case of high level densities and high $Q$ values (Q$\geq$5 MeV), reaction rates can be calculated by statistical models \cite{HGW95}. For nuclei close to proton drip line $Q$ values of (p,$\gamma$) reactions drop strongly, the compound nuclei are produced at low densities \cite{WoG94}. Proton capture reaction rates for low level densities in the compound nucleus are determined by contributions of direct capture and single resonances correlated with unbound states in the compound nucleus. 

\subsection{Direct capture}
 
Nonresonant cross section is expressed in terms of the astrophysical $S$--factor, which contains the nuclear component of the transition probability to the ground states and low excited states of final nuclei. $S$--factors are calculated using RADCAP code \cite{bert03}, which uses a Woods--Saxon nuclear potential (central + spin orbit) and a Coulomb potential of a uniform charge distribution. The nuclear potential parameters are chosen to reproduce the bound state energies. 

Direct proton capture rate on the target nucleus in state $i$ is given by
\begin{align} 
\label{DCint}
& N_A<\sigma v>_{\rm{dc}\, i} = N_A{\left(\frac{8}{\pi \mu}\right)}^{1/2}\frac{1}{\left(k T\right)^{3/2}} \nonumber\\
& \times \int^{\infty}_{0}{S_i(E)\exp\left[-\frac{E}{k T}-\frac{b}{E^{1/2}}\right]dE} ,
\end{align}
where b, which arises from the barrier penetrability, is given by
\begin{equation}
b = {\left(2 \mu\right)}^{1/2}\pi e^2 Z_1 Z_2/\hbar = 0.989 Z_1 Z_2 {\mu}^{1/2}\ ( MeV)^{1/2}.
\end{equation}
Here $\mu$ is reduced mass in entrance channel in amu, $k$ is the Boltzmann constant, $Z$ is a charge number of target nucleus, and $E$ is proton energy. $S_i$ is the sum of the individual $S$--factors of the transitions from the initial state $i$ into all bound states in the final nucleus.
For a given stellar temperature $T$, nuclear reactions take place in relatively narrow energy window around the effective burning energy $E_0$. The $S$--factor is nearly a constant over Gamov Window $E_0 \pm \Delta/2$, where
\begin{equation}
E_0 = 1.22{\left(Z^{2}_{j} Z^{2}_{k} A T^{2}_{6}\right)}^{1/3}\ (keV)
\end{equation}
and
\begin{equation}
\Delta = 0.749{\left(Z^{2}_{j} Z^{2}_{k} A T^{5}_{6}\right)}^{1/6}\ (keV) .
\end{equation}
Hence, the capture rate $N_A<\sigma v>_{{\rm dc}\, i}$ is usually approximated by a constant $S$--factor. 

\subsection{Resonant capture}

Resonance decay widths depend on the Coulomb barrier, the orbital momentum barrier and partial widths of the compound state. The resonant reaction rate for capture on a nucleus in an initial state $i$,
$N_A <\sigma v>_{ {\rm res}\, i}$ for isolated narrow resonances is calculated 
as a sum over all relevant compound nucleus states $j$ above the proton threshold \cite{FoH64}:

\begin{align} 
\label{EqRes}
& N_A <\sigma v>_{{\rm res}\, i} = 1.540 \times 10^{11} (\mu T_9)^{-3/2} \nonumber\\
& \times \sum_j \omega\gamma_{ij} {\rm e}^{-E_{ij}/(kT)} \, \, \, {\rm cm^3 \, s^{-1} mole^{-1}}
\end{align}

Here $T_9$ is the temperature in $ GK$, $E_{ij} = E_j - Q - E_i$ is the resonance energy in the center of mass system, the resonance strengths in $ MeV$ for proton capture are

\begin{equation}
\omega \gamma_{ij} = \frac{2J_j+1}{2(2J_i+1)} \frac{\Gamma_{{\rm p}\, ij} \Gamma_{\gamma j}}
{\Gamma_{{\rm total}\, j}}
\end{equation}

$\Gamma_{{\rm total}\, j}$ is a total width of the resonance level and $J_i$ and $J_j$ are target and projectile spins. 
Proton decay width depends exponentially on the resonance energy and can be calculated from proton spectroscopic factor $C^2S_{ij}$ and the single particle proton width 
$\Gamma_{{\rm sp} \, ij}$ as $\Gamma_{{\rm p} \, ij} = C^2S_{ij} \Gamma_{{\rm sp} \,ij}$. The single particle proton widths were calculated from $\Gamma_{{\rm sp} }=2 \gamma^{2}P\left(l,R_{c}\right)$ \cite{LaTh58} with $\gamma^{2}=\frac{\hbar^{2}c^{2}}{2\mu R^{2}_{c}}$ and where the channel radius $R_{c}$ was chosen to match the width obtained from an exact evaluation of the proton scattering cross section from a Woods-Saxon potential well for $A=32$, $Q=0.1-0.5$  MeV. This simple model matches exact calculations in the sd-shell t within about 10\%, and has advantage that it is fast and can be easily extrapolated to energies below 0.1  MeV where the scattering calculation becomes computationally difficult. We use a Coulomb penetration code from Barker \cite{Ba01}.

The proton spectroscopic factors and the gamma widths are obtained with the  USD interaction \cite{B88} in the full sd-shell model space. The gamma decay widths were obtained with B(M1) values based on free-nucleon g-factors and B(E2) values with harmonic oscillator radial wave functions and effective charges of $e_{p}=1.35$ and $e_{n}=0.35$ \cite{B88}. The uncertainty of the spectroscopic factors and gamma decay widths is on the order of  20$\% $. As we will emphasize the largest uncertainty for the present work is the precise location of the resonances near the proton decay thresholds.

\subsection{Total reaction rate}

The total reaction rate is the sum of the capture rate on all thermally excited states in the target nucleus weighted with their individual population factors:

\begin{align}
 N_A <\sigma v> = & \sum_i (N_A<\sigma v>_{{\rm res} \, i} +  N_A<\sigma v>_{{\rm dc}\, i} ) \nonumber\\
& \times \frac{(2J_i+1) {\rm e}^{-E_i/kT}}  {\sum_n (2J_n+1) {\rm e}^{-E_n/kT}}.
\end{align}

Here $E_i$ and $E_n$ are the resonance energies.

\section{Results}

\subsection{$^{32}$Cl(p,$\gamma$)$^{33}$Ar}

\begin{table*}
\caption{\label{32ClDC} Spectroscopic factors $C^2S$ and astrophysical S-factors $S(E_0)$ for direct capture into bound states in $^{33}$Ar for temperature range of $0.1-2.0$ GK. Listed are results for capture on the $^{32}$Cl ground state, $J^\pi=1^+$, as well as on the first excited state in  $^{32}$Cl, $J^\pi=2^+$ (denoted with an asterix). $J^\pi$ are spin and parity of the $^{33}$Ar final state, $n$ is the node number, $l_0$ the single particle orbital momentum, and $j_0$ the total single particle angular momentum. $E_x$ is excitation energy.}
\begin{ruledtabular}
\begin{tabular}{dccdcdc}
 E_x {\rm ( MeV)} & $J^\pi$ & $(nl_0)_{j0}$ & C^2S & $S(E_0) {\rm ( MeV barn)}$ &  C^2S^* & $S(E_0)^* {\rm ( MeV barn)}$ \\
0     & $1/2^+$ & 2s$_{1/2}$   & 0.080  & 6.58e-3 -- 7.00e-3     &           &                    \\
      &         & 1d$_{3/2}$   & 0.672  & 3.78e-3 -- 4.43e-3     & 1.127     & 4.15e-3 -- 5.32e-3 \\
      &         & 1d$_{5/2}$   &        &                        & 0.001     & 4.09e-6 -- 6.38e-6 \\
1.359 & $3/2^+$ & 2s$_{1/2}$   & 0.001  & 8.65e-5 -- 9.63e-5     & 0.006     & 5.76e-4 -- 6.19e-4 \\
      &         & 1d$_{3/2}$   & 0.185  & 1.54e-3 -- 1.82e-3     & 0.119     & 6.35e-4 -- 8.17e-4 \\
      &         & 1d$_{5/2}$   & 0.004  & 5.74e-5 -- 7.76e-5     & 0.009     & 7.73e-5 -- 1.20e-4 \\ 
1.798 & $5/2^+$ & 2s$_{1/2}$   &        &                        & 0.002     & 2.42e-4 -- 2.69e-4 \\
      &         & 1d$_{3/2}$   & 0.145  & 1.60e-3 -- 1.91e-3     & 0.620     & 4.33e-3 -- 5.60e-3 \\
      &         & 1d$_{5/2}$   & 0.006  & 1.07e-4 -- 1.46e-4     & 0.021     & 2.25e-4 -- 3.49e-4 \\
2.439 & $3/2^+$ & 2s$_{1/2}$   & 0.031  & 3.68e-3 -- 5.20e-3     & 0.024     & 1.83e-3 -- 2.36e-3 \\
      &         & 1d$_{3/2}$   & 0.167  & 9.65e-4 -- 1.18e-3     & 0.128     & 4.68e-4 -- 6.18e-4 \\
      &         & 1d$_{5/2}$   & 0.014  & 1.19e-4 -- 1.66e-4     & 0.016     & 8.99e-5 -- 1.42e-4 \\
3.154 & $3/2^+$ & 2s$_{1/2}$   & 0.068  & 6.92e-3 -- 1.53e-2     & 0.001     & 4.01e-5 -- 7.36e-5 \\
      &         & 1d$_{3/2}$   & 0.516  & 1.92e-3 -- 2.67e-3     & 0.169     & 4.11e-4 -- 5.91e-4 \\
      &         & 1d$_{5/2}$   & 0.003  & 1.55e-5 -- 2.40e-5     &           &                    \\ \hline
\mbox{total} &    &            &        & 2.95e-2 -- 3.74e-2     &           & 1.37e-2 -- 1.63e-2 \\
\end{tabular}
\end{ruledtabular}
\end{table*}

\begin{table*}[htbp] 
\caption{\label{32Cl}Properties of resonant states for $^{32}$Cl$(p,\gamma)$ reaction. Listed are spin and parity $J^\pi$, excitation energy $E_x$, center of mass resonance energy $E_r$, proton single particle widths $\Gamma_{\rm sp}$ for angular momenta $l$, spectroscopic factors $C^2S$, proton-decay width $\Gamma_p$, $\gamma$-decay width $\Gamma_\gamma$ and the
resonance strength $\omega\gamma$. The upper part is for ground state capture, $J^\pi=1^+$, the lower part for capture on the first excited state in $^{32}$Cl, $J^\pi=2^+$.}
\begin{ruledtabular}
\begin{tabular}{cccccccccc}
$J^\pi$  & $E_x {\rm ( MeV)}$ & $ E_r {\rm ( MeV)}$ & \multicolumn{2}{c}{$\Gamma_{\rm sp}$}  & \multicolumn{2}{c}{$C^2S$} & $\Gamma_{\rm p} {\rm (eV)}$ & $\Gamma_\gamma {\rm (eV)}$ & $\omega\gamma {\rm (eV)}$ \\
 & & & l=0 & l=2 & l=0 & l=2 \\  
$5/2^{+}_{2}$ & 3.364 & 0.021 $\pm$ 0.09 &            &  8.000e-42 &           &  2.900e-02 &  2.320e-43 &  2.638e-02 &  2.320e-43 \\ 
$7/2^{+}_{1}$ & 3.456 & 0.113 $\pm$ 0.09 &            &  1.700e-13 &           &  2.860e-03 &  4.862e-16 &  3.343e-03 &  6.483e-16 \\ 
$5/2^{+}_{3}$ & 3.819 & 0.476 $\pm$ 0.008 &           &  2.240e-02 &           &  4.226e-02 &  9.466e-04 &  1.500e-02 &  7.144e-04 \\ 
$1/2^{+}_{2}$ & 4.190 & 0.847 $\pm$ 0.100 &  6.400e+02 &  1.060e+01 &  6.703e-02 &  3.384e-02 &  4.326e+01 &  2.260e-01 &  7.492e-02 \\ 
$3/2^{+}_{4}$ & 4.730 & 1.387 $\pm$ 0.100 &  2.280e+04 &  5.874e+02 &  2.890e-03 &  3.904e-02 &  8.882e+01 &  1.133e-01 &  5.527e-03 \\ 
\hline\hline
$7/2^{+}_{1}$ & 3.456 & 0.023 $\pm$ 0.09 &            &  1.300e-39 &           &  4.650e-03 &  6.045e-42 &  3.343e-03 &  4.836e-42 \\
$5/2^{+}_{3}$ & 3.819 & 0.386 $\pm$ 0.008 &  1.340e-01 &  1.500e-03 &  2.440e-02 &  4.387e-01 &  3.928e-03 &  1.500e-02 &  1.779e-03 \\ 
$1/2^{+}_{2}$ & 4.190 & 0.757 $\pm$ 0.100 &           &  3.400e+00 &           &  2.260e-03 &  7.684e-03 &  2.260e-01 &  7.985e-06 \\ 
$3/2^{+}_{4}$ & 4.730 & 1.297 $\pm$ 0.100 &  1.500e+04 &  3.713e+02 &  7.491e-02 &  3.780e-03 &  1.125e+03 &  1.133e-01 &  4.200e-02 \\
\end{tabular}
\end{ruledtabular}
\end{table*}

\begin{figure*}[htbp]
\includegraphics[width=8cm]{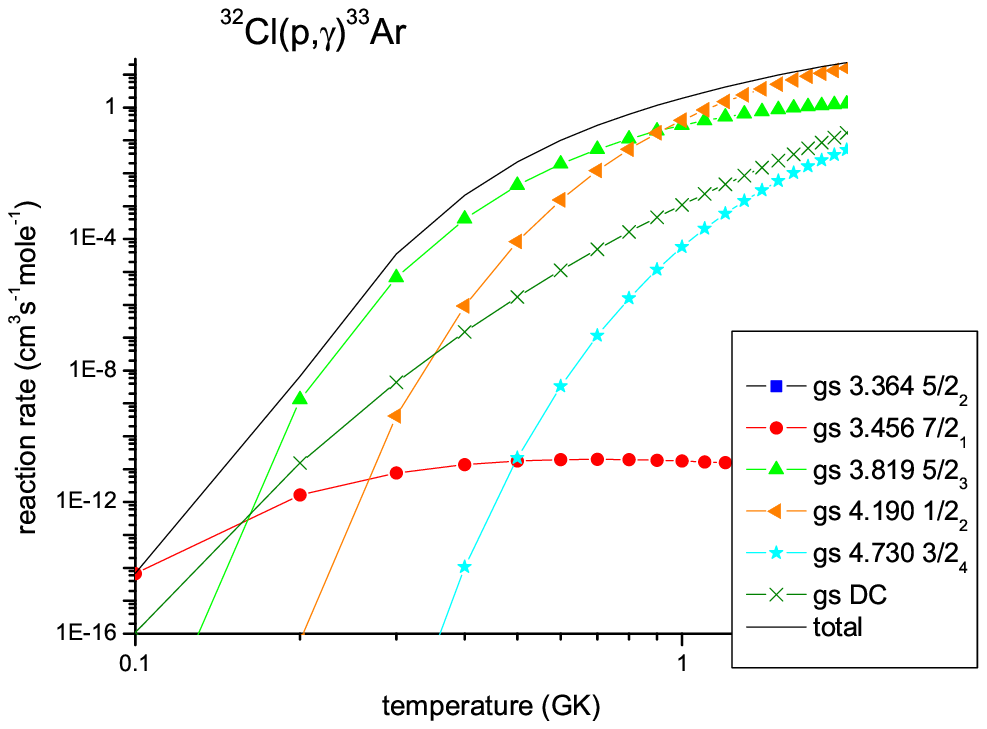}
\includegraphics[width=8cm]{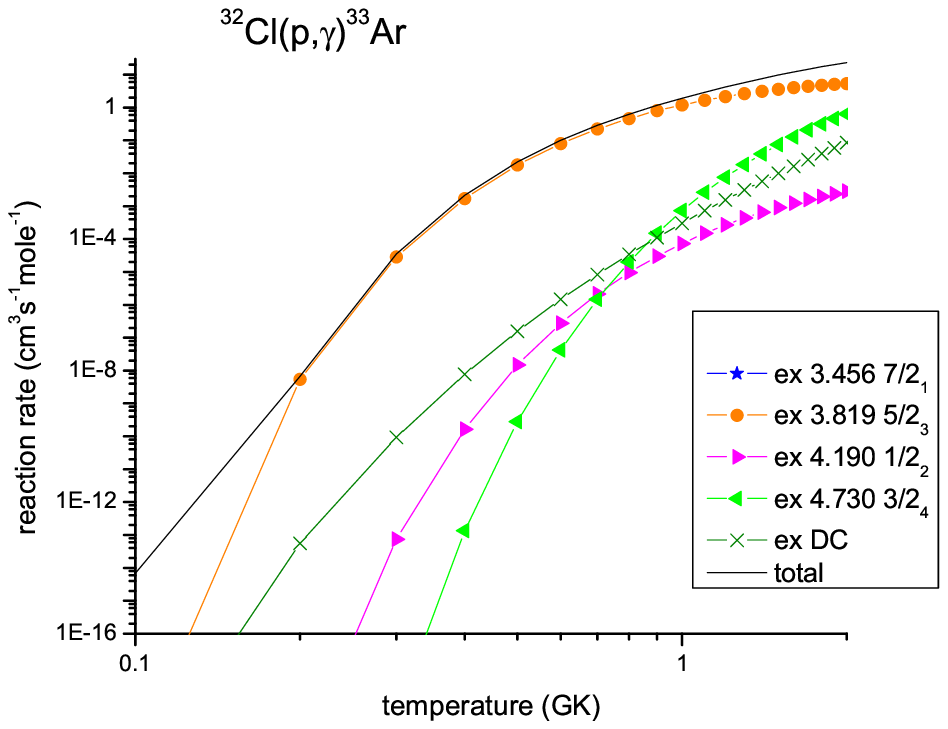}
\caption{\label{32res} The contributions of various individual resonances and through direct capture (DC) to the $^{32}$Cl(p,$\gamma$)$^{33}$Ar reaction rate as functions of temperature. In the legend, resonances are labeled with their excitation energy in $^{33}$Ar. The right panel shows contributions from capture on the ground state in $^{32}$Cl, $J^\pi=1^+$, the left panel contributions from capture on the first excited state, $J^\pi=2^+$, in each case weighted with relative population of the respective target state. Both panels show the same total $^{32}$Cl(p,$\gamma$)$^{33}$Ar reaction rate for comparison.}
\end{figure*}

We start with the application to the previously studies case of $^{32}$Cl(p,$\gamma$)$^{33}$Ar. As an odd-odd nucleus, $T=1$, $^{32}$Cl has a low lying excited state at 89.9keV. Based on it's mirror nucleus $^{32}$P the spin of the state is $2^+$. The ground state spin is $1^+$. Proton separation energy of $^{33}$Ar is 3.343  MeV. As shown in Fig.~\ref{32res}, the proton capture rate is dominant by resonant capture on the first excited state, as it was shown in Schatz et al. \cite{Sch05}. Gamow window for $T=0.1T_9$ is at 127.5--216.5 keV, and for $T=2T_9$ is at 727.5--1807.4 keV. Tab.~\ref{32ClDC} shows astrophysical S--factor range for these temperatures.
 
\begin{figure}[htbp]
\includegraphics[width=8cm]{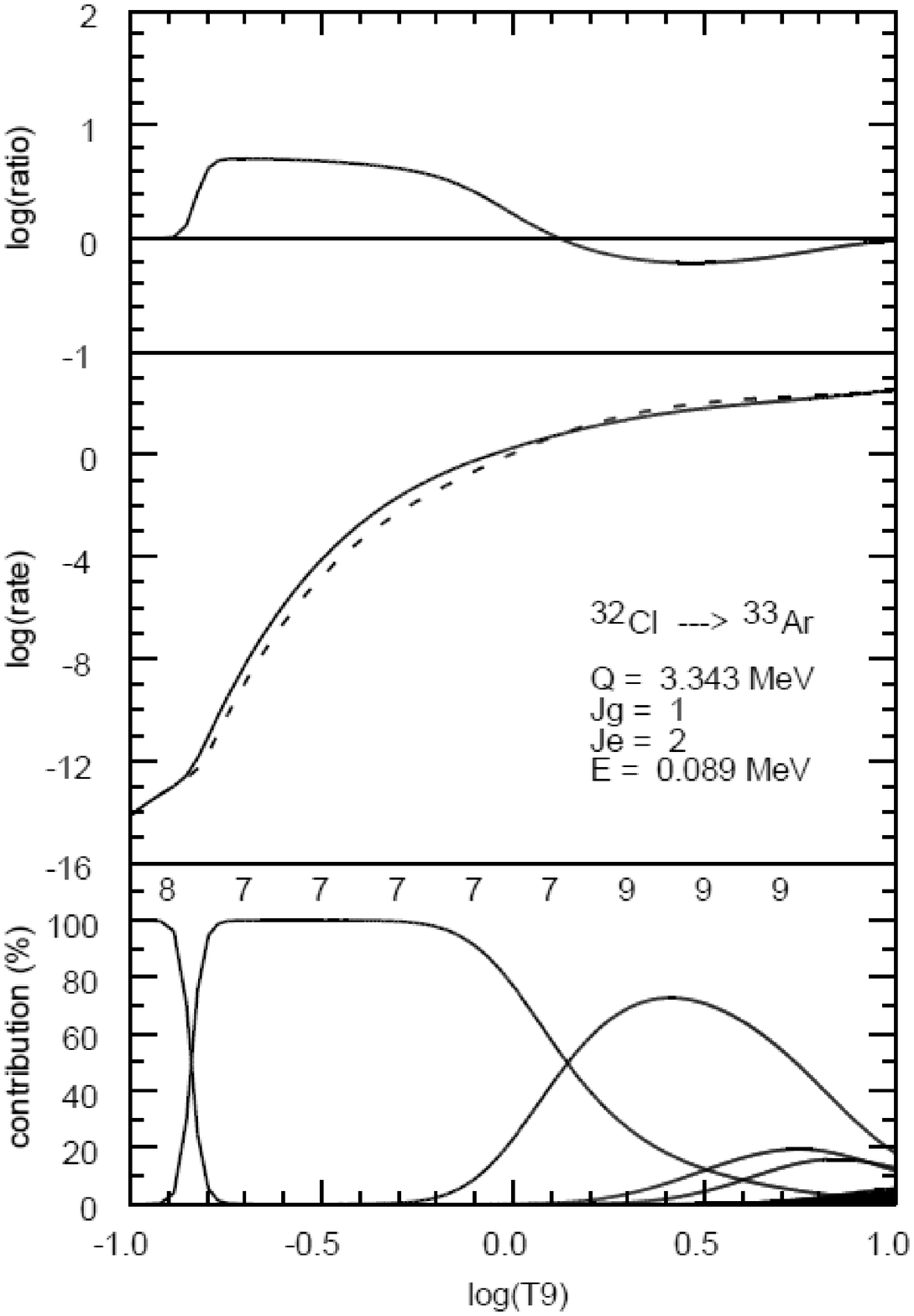}
\caption{\label{Fig1} Results for $^{32}$Cl$(1^{+},2^{+}) \rightarrow ^{33}$Ar. The middle panel shows the rp-reaction rate for the $^{32}$Cl $1^{+}$ ground state (dashed line) and the sum of the contributions from the ground and a $2^{+}$ excited state at 0.089  MeV (full line). The uper panel shows the stellar enhancement factor - the ratio of the total rate over the ground state reaction rate. The bottom panel shows the percentage contribution of the total rate from each of the final states with number $f$  of the most important indicated in the top of the bottom panel.}
\end{figure}

Properties of the resonance states in $^{33}$Ar are listed in Tab.~\ref{32Cl} and properties of the bound states are listed in Tab.~\ref{32ClDC}. Single particle proton widths for second $5/2$ state for proton capture on a ground state and for first $7/2$ state for proton capture on a first excited state were taken from Schatz et al. \cite{Sch05}.  Calculations were done using experimental resonance energies \cite{CBB04}, gamma widths were obtained using shell model energies. The USD energies (experimental excitation energies shown in brackets) for the states in $^{33}$Ar near proton decay threshold are 3.606 (3.154)  MeV for (5,3/2$^{+}$,3), 3.787 (3.364)  MeV for (6,5/2$^{+}$,2), 3.949 (3.456)  MeV for (8,7/2$^{+}$,1), 3.888 (3.819)  MeV for (7,5/2$^{+}$,3), 4.242 (4.190)  MeV for (9,1/2$^{+}$,2) and  4.997 (4.730)  MeV for (10,3/2$^{+}$,4). The states are labeled by ($f,J^{\pi},k$), where $f$ is the level number obtained with the USD Hamiltonian ad $k$ is the level number for a given $J^{\pi}$ value. The effect of the energy shift on the gamma decay width is small (20\% or less) and is not taken into account. The dashed line in the middle panel of Fig. \ref{Fig1} shows the reaction rate when only the $^{32}$Cl ground state is considered. The full line in the middle panel shows the total tate with both the ground state and first excited state of  $^{32}$Cl are cosidered. The ratio of the total rate to the ground state rate is the stellar-enhancement factor (SEF) shown in the top panel in Fig. \ref{Fig1}. The bottom panel in Fig. \ref{Fig1} shows the contribution from each final state in $^{33}$Ar with the number $f$ for the largest of each labeled along the top. It is noted that state number $f=7$ is most important over the temperature range 0.1 to 0.8  GK. The large SEF is due to the $l=0$ spectroscopic factor. We will find in general that large SEF are associated with $l=0$ spectroscopic factors.

Are other states important? The next state in $^{32}$Cl is a $0^{+}$ state at 0.466  MeV. The reaction rate contribution for this state is negligible compared to the lower states, in the temperature range we consider the largest SEF is 0.90 at 10  GK. For an energy of 0.461  MeV the factor of 
\begin{equation}
e^{-E_{i}/(kT)}=  e^{-11.60( GK/ MeV)E_{i}/T}
\end{equation}
is small. And in addition there are no final states that enter with very large spectroscopic factor to this excited state. Thus we only need to consider initial nuclei with excited states below about 300  keV. 

\subsection{$^{31}$Cl(p,$\gamma$)$^{32}$Ar}

\begin{table*}
\caption{\label{31ClDC} Spectroscopic factors $C^2S$ and astrophysical S-factors $S(E_0)$ for direct capture into bound states in $^{32}$Ar for temperature range of $0.1-2.0$ GK. Listed are results for capture on the $^{32}$Cl ground state, $J^\pi=3/2^+$, as well as on the first excited state in  $^{31}$Cl, $J^\pi=1/2^+$ (denoted with an asterix). $J^\pi$ are spin and parity of the $^{32}$Ar final state, $n$ is the node number, $l_0$ the single particle orbital momentum, and $j_0$ the total single particle angular momentum. $E_x$ is excitation energy.}
\begin{ruledtabular}
\begin{tabular}{dccdcdc}
 E_x {\rm ( MeV)} & $J^\pi$ & $(nl_0)_{j0}$ & C^2S & $S(E_0) {\rm ( MeV barn)}$ &  C^2S^* & $S(E_0)^* {\rm ( MeV barn)}$ \\
0     & $0^+$   & 2s$_{1/2}$   &        &                        & 1.201     & 1.70e-1 -- 2.25e-1 \\
      &         & 1d$_{3/2}$   & 1.905  & 8.52e-3 -- 9.82e-3     &           &                    \\
2.093 & $2^+$   & 2s$_{1/2}$   & 0.010  & 1.80e-3 -- 3.25e-3     &           &                    \\
      &         & 1d$_{3/2}$   & 0.637  & 6.23e-3 -- 7.96e-3     & 0.190     & 5.69e-3 -- 7.28e-3 \\
      &         & 1d$_{5/2}$   & 0.056  & 8.77e-4 -- 1.35e-3     & 0.042     & 2.08e-3 -- 3.47e-3 \\ \hline
\mbox{total} &    &            &        & 1.83e-2 -- 2.10e-2     &           & 1.78e-1 -- 2.36e-1 \\
\end{tabular}
\end{ruledtabular}
\end{table*}

\begin{table*}[htbp] 
\caption{\label{31Cl}Properties of resonant states for $^{31}$Cl$(p,\gamma)$ reaction. Listed are spin and parity $J^\pi$, excitation energy $E_x$, center of mass resonance energy $E_r$, proton single particle widths $\Gamma_{\rm sp}$ for angular momenta $l$, spectroscopic factors $C^2S$, proton-decay width $\Gamma_p$, $\gamma$-decay width $\Gamma_\gamma$ and the
resonance strength $\omega\gamma$. The upper part is for ground state capture, $J^\pi=3/2^+$, the lower part for capture on the first excited state in $^{31}$Cl, $J^\pi=1/2^+$.}
\begin{ruledtabular}
\begin{tabular}{cccccccccc}
$J^\pi$  & $E_x {\rm ( MeV)}$ & $ E_r {\rm ( MeV)}$ & \multicolumn{2}{c}{$\Gamma_{\rm sp}$}  & \multicolumn{2}{c}{$C^2S$} & $\Gamma_{\rm p} {\rm (eV)}$ & $\Gamma_\gamma {\rm (eV)}$ & $\omega\gamma {\rm (eV)}$ \\
 & & & l=0 & l=2 & l=0 & l=2 \\  
$2^{+}_{2}$ & 4.212 & 1.790  &  7.350e+04 &  3.400e+03 &  6.246e-02 &  6.455e-01 &  6.785e+03 &  5.410e-04 &  3.379e-04 \\ 
$0^{+}_{2}$ & 4.808 & 2.386  &            &  1.950e+04 &            &  6.100e-03 &  1.190e+02 &  3.531e-04 &  3.712e-07 \\ 
$1^{+}_{1}$ & 5.581 & 3.159  &            &  8.540e+04 &  1.478e-01 &  8.105e-02 &  6.922e+03 &  1.415e-01 &  2.900e-02 \\ 
\hline\hline
$2^{+}_{2}$ & 4.212 & 1.043  &           &  5.950e+01 &           &  5.963e-02 &  3.548e+00 &  5.410e-04 &  3.534e-07 \\ 
$0^{+}_{2}$ & 4.808 & 1.639  & 5.810e+04 &            & 2.414e-01 &            &  1.402e+04 &  3.531e-04 &  8.754e-05 \\ 
$1^{+}_{1}$ & 5.581 & 2.412  & 1.017e+05 &  2.050e+04 &  7.000e-04 &  2.765e-01 &  5.739e+03 &  1.415e-01 &  4.809e-02 \\
\end{tabular}
\end{ruledtabular}
\end{table*}

The first excited state of $^{31}$Cl, spin $1/2^+$, is at 747 keV. Spin of the ground state is $3/2^+$. The second excited state, spin $5/2^+$, is at 1292 keV. USD shell model gives energies of 815 keV and 1606 keV. Proton separation energy of $^{32}$Ar is 2.42 MeV, $T=3/2$. Since excitation energies of $^{32}$Ar are unknown, USD shell model predictions were used.

\begin{figure*}[htbp]
\includegraphics[width=8cm]{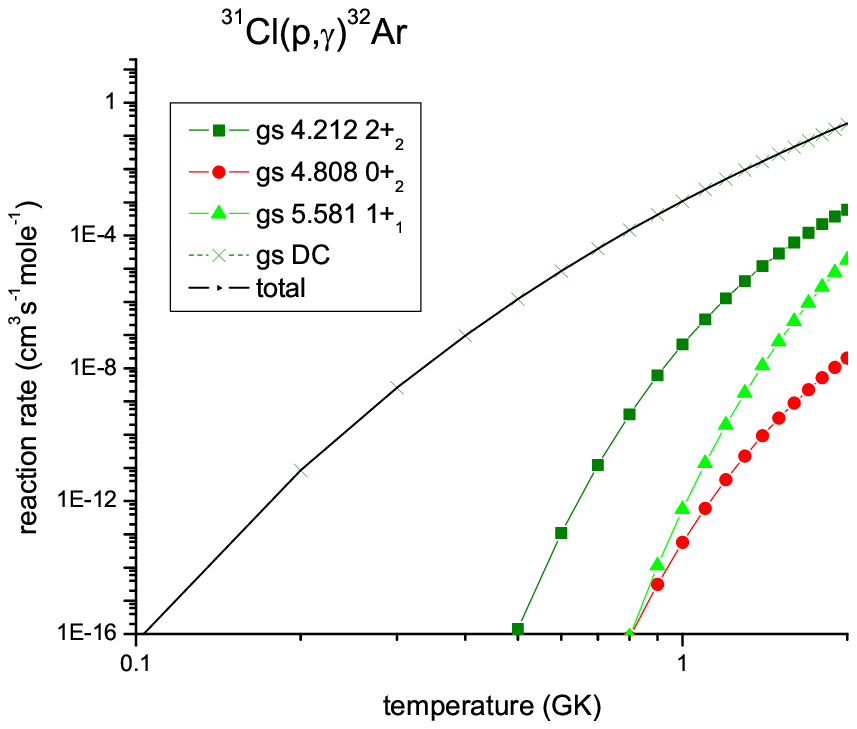}
\includegraphics[width=8cm]{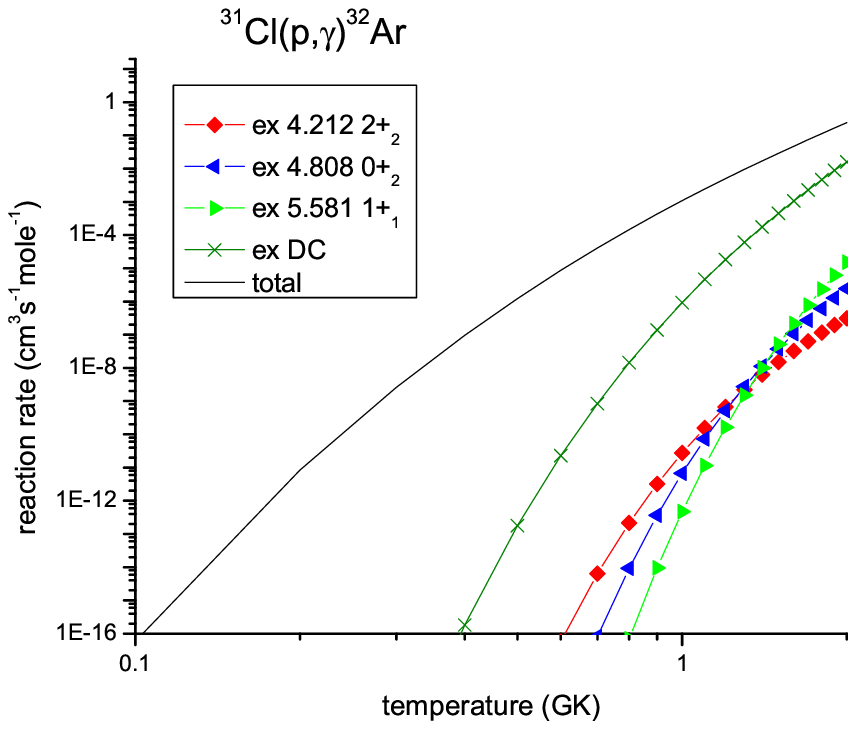}
\caption{\label{31res} The contributions of various individual resonances and through direct capture (DC) to the $^{31}$Cl(p,$\gamma$)$^{32}$Ar reaction rate as functions of temperature. In the legend, resonances are labeled with their excitation energy in $^{32}$Ar. The right panel shows contributions from capture on the ground state in $^{31}$Cl, $J^\pi=3/2^+$, the left panel contributions from capture on the first excited state, $J^\pi=1/2^+$, in each case weighted with relative population of the respective target state. Both panels show the same total $^{31}$Cl(p,$\gamma$)$^{32}$Ar reaction rate for comparison.}
\end{figure*}

According to USD shell model, $^{32}$Ar has a first excited state at 2.093 MeV, therefore proton capture is dominated by direct capture. As seen from Fig. ~\ref{31res} direct capture only on a ground state of $^{31}$Cl is important.

\subsection{$^{23}$Al(p,$\gamma$)$^{24}$Si}

\begin{table*}
\caption{\label{23AlDC} Spectroscopic factors $C^2S$ and astrophysical S-factors $S(E_0)$ for direct capture into bound states in $^{24}$Si for temperature range of $0.1-2.0 GK$. Listed are results for capture on the $^{23}$Al ground state, $J^\pi=5/2^+$, as well as on the first excited state in  $^{23}$Al, $J^\pi=1/2^+$ (denoted with an asterix). $J^\pi$ are spin and parity of the $^{24}$Si final state, $n$ is the node number, $l_0$ the single particle orbital momentum, and $j_0$ the total single particle angular momentum. $E_x$ is excitation energy.}
\begin{ruledtabular}
\begin{tabular}{dccdcdc}
 E_x {\rm ( MeV)} & $J^\pi$ & $(nl_0)_{j0}$ & C^2S & $S(E_0) {\rm ( MeV barn)}$ &  C^2S^* & $S(E_0)^* {\rm ( MeV barn)}$ \\
0     & $0^+$   & 2s$_{1/2}$   &        &                        & 0.226     & 5.42e-3 -- 8.11e-3 \\
      &         & 1d$_{5/2}$   & 3.461  & 2.52e-3 -- 2.82e-3     &           &                    \\
2.145 & $2^+$   & 2s$_{1/2}$   & 0.191  & 3.90e-3 -- 5.31e-3     &           &                    \\
      &         & 1d$_{3/2}$   & 0.025  & 3.13e-5 -- 3.33e-5     & 0.009     & 4.50e-5 -- 6.37e-5 \\
      &         & 1d$_{5/2}$   & 0.204  & 4.12e-4 -- 4.82e-4     & 0.043     & 3.69e-4 -- 6.77e-4 \\ \hline
\mbox{total} &    &            &        & 7.24e-3 -- 8.27e-3     &           & 5.83e-3 -- 8.85e-3 \\
\end{tabular}
\end{ruledtabular}
\end{table*}

\begin{table*}[htbp] 
\caption{\label{23Al}Properties of resonant states for $^{23}$Al$(p,\gamma)$ reaction. Listed are spin and parity $J^\pi$, excitation energy $E_x$, center of mass resonance energy $E_r$, proton single particle widths $\Gamma_{\rm sp}$ for angular momenta $l$, spectroscopic factors $C^2S$, proton-decay width $\Gamma_p$, $\gamma$-decay width $\Gamma_\gamma$ and the
resonance strength $\omega\gamma$. The upper part is for ground state capture, $J^\pi=5/2^+$, the lower part for capture on the first excited state in $^{23}$Al, $J^\pi=1/2^+$.}
\begin{ruledtabular}
\begin{tabular}{cccccccccc}
$J^\pi$  & $E_x {\rm ( MeV)}$ & $ E_r {\rm ( MeV)}$ & \multicolumn{2}{c}{$\Gamma_{\rm sp}$}  & \multicolumn{2}{c}{$C^2S$} & $\Gamma_{\rm p} {\rm (eV)}$ & $\Gamma_\gamma {\rm (eV)}$ & $\omega\gamma {\rm (eV)}$ \\
 & & & l=0 & l=2 & l=0 & l=2 \\  
$2^{+}_{2}$ & 3.742 & 0.442  &  6.450e+01 &  5.808e-01 &  5.000e-01 &  1.403e-01 &  3.233e+01 &  1.410e-02 &  5.873e-03 \\
$4^{+}_{1}$ & 3.996 & 0.696  &            &  2.200e+01 &            &  1.314e-02 &  2.891e-01 &  3.979e-04 &  2.980e-04 \\ 
$3^{+}_{1}$ & 4.568 & 1.268  &  7.370e+04 &  1.000e+03 &  4.912e-01 &  2.257e-01 &  3.663e+04 &  7.329e-03 &  4.272e-03 \\ 
$0^{+}_{2}$ & 4.662 & 1.362  &            &  3.500e+03 &            &  2.025e-01 &  7.088e+02 &  1.618e-04 &  6.463e-07 \\ 
$2^{+}_{3}$ & 5.339 & 2.039  &  1.325e+05 &  2.420e+04 &  3.386e-02 &  4.454e-02 &  5.564e+03 &  1.661e-01 &  6.677e-02 \\ 
$3^{+}_{2}$ & 5.473 & 2.173  &            &  3.270e+04 &  2.390e-02 &  1.348e-01 &  4.408e+03 &  1.607e-02 &  9.027e-03 \\ 
\hline\hline
$3^{+}_{1}$ & 4.568 & 0.808  &           &  1.085e+02 &           &  2.779e-01 &  3.015e+01 &  7.329e-03 &  1.055e-05 \\ 
$0^{+}_{2}$ & 4.662 & 0.902  & 1.300e+04 &            & 1.083e+00 &            &  1.408e+04 &  1.618e-04 &  3.852e-05 \\ 
$2^{+}_{3}$ & 5.339 & 1.579  &           &  6.300e+03 &           &  3.201e-02 &  2.017e+02 &  1.661e-01 &  7.259e-03 \\
$3^{+}_{2}$ & 5.473 & 1.713  &           &  1.230e+04 &           &  1.390e-02 &  1.710e+02 &  1.607e-02 &  1.050e-03 \\
\end{tabular}
\end{ruledtabular}
\end{table*}

ENSDF site gives the first excited state of $^{23}$Al at 460 keV, but it's spin is $3/2^+$, which is inconsistent with it's mirror nucleus $^{23}$Ne.  $^{23}$Ne has a ground state spin $5/2^+$ and it's $3/2^+$ state is at 1822 keV. Therefore calculations were done considering the first excited state at 460 keV with spin $1/2^+$. Shell model gives the first excited state of $^{23}$Al at 992 keV. Proton separation energy of $^{24}$Si is 3.3 MeV, $T=3/2$. Experimental excitation energies of $^{24}$Si are unknown, therefore Shell model energy predictions were used.

\begin{figure*}[htbp]
\includegraphics[width=8cm]{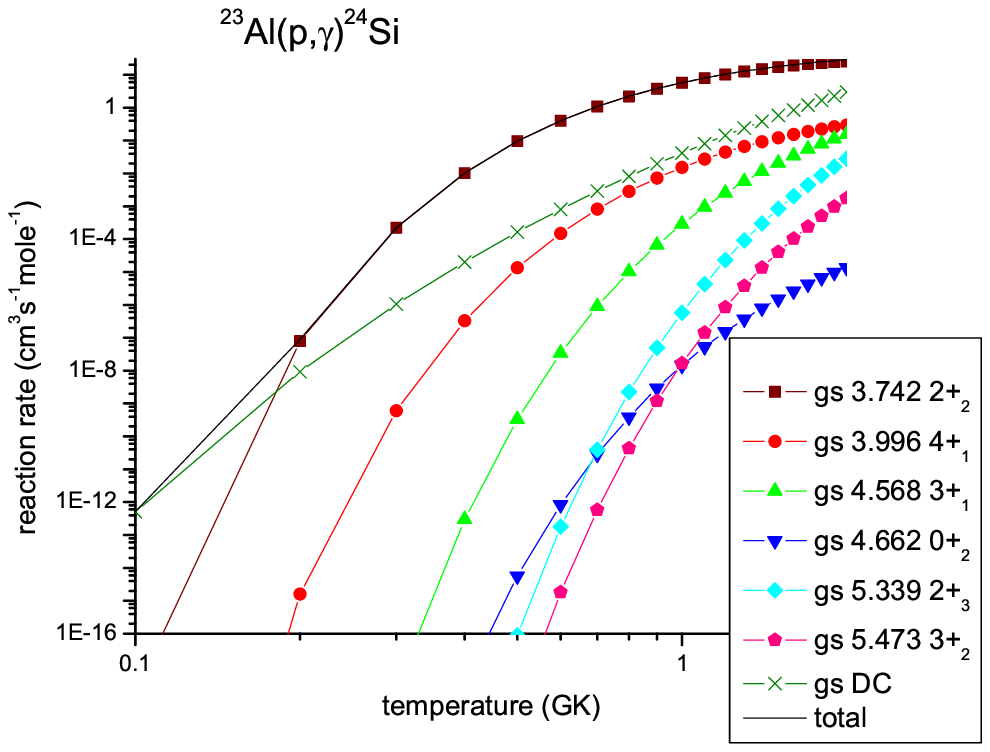}
\includegraphics[width=8cm]{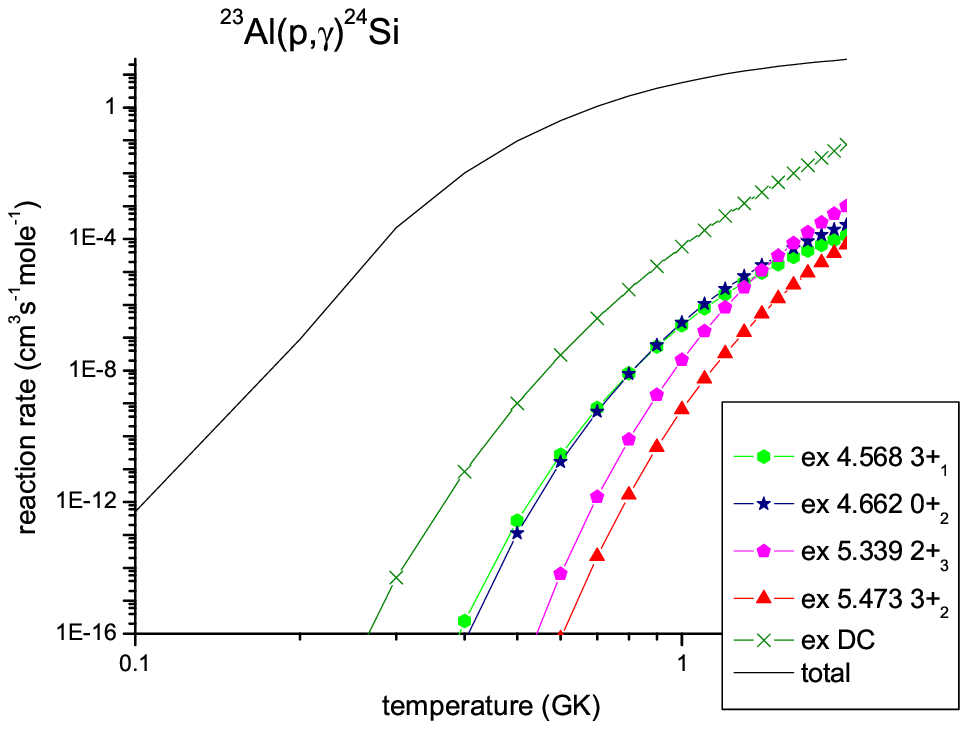}
\caption{\label{23res} The contributions of various individual resonances and through direct capture (DC) to the $^{23}$Al(p,$\gamma$)$^{24}$Si reaction rate as functions of temperature. In the legend, resonances are labeled with their excitation energy in $^{24}$Si. The right panel shows contributions from capture on the ground state in $^{23}$Al, $J^\pi=5/2^+$, the left panel contributions from capture on the first excited state, $J^\pi=1/2^+$, in each case weighted with relative population of the respective target state. Both panels show the same total $^{23}$Al(p,$\gamma$)$^{24}$Si reaction rate for comparison.}
\end{figure*}

As seen from Fig.~\ref{23res} total reaction rate is dominated by resonant capture on the ground state of $^{23}$Al via the second $2^+$ state resonance of 3.742 MeV (USD shell model prediction), which agrees with Herndl et al. calculations \cite{HGW95}.

\subsection{$^{24}$Al(p,$\gamma$)$^{25}$Si}

\begin{table*}
\caption{\label{24AlDC} Spectroscopic factors $C^2S$ and astrophysical S-factors $S(E_0)$ for direct capture into bound states in $^{25}$Si for temperature range of $0.1-2.0 GK$. Listed are results for capture on the $^{24}$Al ground state, $J^\pi=4^+$ as well as on the first excited state in  $^{24}$Al, $J^\pi=1^+$ (denoted with an asterix). $J^\pi$ are spin and parity of the $^{25}$Si final state, $n$ is the node number, $l_0$ the single particle orbital momentum, and $j_0$ the total single particle angular momentum. $E_x$ is excitation energy.}
\begin{ruledtabular}
\begin{tabular}{dccdcdc}
 E_x {\rm ( MeV)} & $J^\pi$ & $(nl_0)_{j0}$ & C^2S & $S(E_0) {\rm ( MeV barn)}$ &  C^2S^* & $S(E_0)^* {\rm ( MeV barn)}$ \\
0     & $5/2^+$ & 1d$_{3/2}$   & 0.004  & 6.80e-6 -- 6.93e-6     &           &                    \\
      &         & 1d$_{5/2}$   & 0.530  & 1.78e-3 -- 2.01e-3     & 0.070     & 7.82e-4 -- 9.13e-4 \\
0.132 & $3/2^+$ & 2s$_{1/2}$   &        &                        & 0.005     & 2.58e-4 -- 2.63e-4 \\
      &         & 1d$_{3/2}$   &        &                        & 0.001     & 3.07e-6 -- 3.20e-6 \\
      &         & 1d$_{5/2}$   & 0.781  & 1.71e-3 -- 1.93e-3     & 0.483     & 3.53e-3 -- 4.12e-3 \\ 
1.159 & $1/2^+$ & 2s$_{1/2}$   &        &                        & 0.229     & 5.65e-3 -- 5.99e-3 \\
2.130 & $3/2^+$ & 2s$_{1/2}$   &        &                        & 0.030     & 1.25e-3 -- 1.49e-3 \\
      &         & 1d$_{3/2}$   &        &                        & 0.004     & 1.16e-5 -- 1.24e-5 \\
      &         & 1d$_{5/2}$   & 0.006  & 7.84e-6 -- 9.17e-6     &           &                    \\
2.542 & $9/2^+$ & 2s$_{1/2}$   & 0.039  & 1.12e-3 -- 1.66e-3     &           &                    \\
      &         & 1d$_{3/2}$   & 0.039  & 6.48e-5 -- 7.08e-5     &           &                    \\
      &         & 1d$_{5/2}$   & 0.006  & 1.54e-5 -- 1.86e-5     &           &                    \\ 
2.703 & $7/2^+$ & 2s$_{1/2}$   & 0.602  & 1.35e-2 -- 2.13e-2     &           &                    \\
      &         & 1d$_{3/2}$   & 0.004  & 4.84e-6 -- 5.39e-6     &           &                    \\
      &         & 1d$_{5/2}$   & 0.095  & 1.83e-4 -- 2.25e-4     &           &                    \\
2.843 & $5/2^+$ & 1d$_{5/2}$   & 0.008  & 1.02e-5 -- 1.28e-5     & 0.043     & 2.20e-4 -- 2.70e-4 \\
3.019 & $7/2^+$ & 2s$_{1/2}$   & 0.050  & 1.04e-3 -- 1.98e-3     &           &                    \\
      &         & 1d$_{3/2}$   & 0.001  & 9.90e-7 -- 1.18e-6     &           &                    \\
      &         & 1d$_{5/2}$   & 0.005  & 7.85e-6 -- 1.03e-5     & 0.001     & 4.59e-6 -- 5.73e-6 \\ 
3.154 & $9/2^+$ & 2s$_{1/2}$   & 0.329  & 8.28e-3 -- 1.79e-2     &           &                    \\
      &         & 1d$_{3/2}$   & 0.181  & 2.00e-4 -- 2.50e-4     &           &                    \\
      &         & 1d$_{5/2}$   & 0.170  & 2.87e-4 -- 3.94e-4     &           &                    \\
3.304 & $3/2^+$ & 2s$_{1/2}$   &        &                        & 0.280     & 9.22e-3 -- 1.57e-2 \\
      &         & 1d$_{3/2}$   &        &                        & 0.132     & 2.20e-4 -- 2.59e-4 \\
      &         & 1d$_{5/2}$   & 0.002  & 1.17e-6 -- 1.76e-6     & 0.009     & 7.30e-6 -- 1.17e-5 \\ \hline
\mbox{total} &    &            &        & 2.89e-2 -- 4.71e-2     &           & 2.20e-2 -- 2.82e-2 \\
\end{tabular}
\end{ruledtabular}
\end{table*}

\begin{table*}[htbp] 
\caption{\label{24Al}Properties of resonant states for $^{24}$Al$(p,\gamma)$ reaction. Listed are spin and parity $J^\pi$, excitation energy $E_x$, center of mass resonance energy $E_r$, proton single particle widths $\Gamma_{\rm sp}$ for angular momenta $l$, spectroscopic factors $C^2S$, proton-decay width $\Gamma_p$, $\gamma$-decay width $\Gamma_\gamma$ and the
resonance strength $\omega\gamma$. The upper part is for ground state capture, $J^\pi=4^+$, the lower part for capture on the first excited state in $^{24}$Al, $J^\pi=1^+$.}
\begin{ruledtabular}
\begin{tabular}{cccccccccc}
$J^\pi$  & $E_x {\rm ( MeV)}$ & $ E_r {\rm ( MeV)}$ & \multicolumn{2}{c}{$\Gamma_{\rm sp}$}  & \multicolumn{2}{c}{$C^2S$} & $\Gamma_{\rm p} {\rm (eV)}$ & $\Gamma_\gamma {\rm (eV)}$ & $\omega\gamma {\rm (eV)}$ \\
 & & & l=0 & l=2 & l=0 & l=2 \\  
$3/2^{+}_{4}$ & 3.623 & 0.215  &            &  1.248e-04 &           &  9.705e-02 &  1.211e-05 &  4.899e-03 &  2.685e-06 \\ 
$5/2^{+}_{3}$ & 3.657 & 0.249  &            &  9.778e-04 &           &  2.840e-02 &  2.777e-05 &  1.607e-03 &  9.099e-06 \\ 
$9/2^{+}_{3}$ & 3.918 & 0.510  &  2.184e+02 &  2.000e+00 &  2.235e-02 &  1.626e-01 &  5.206e+00 &  1.156e-04 &  6.420e-05 \\ 
$7/2^{+}_{3}$ & 4.003 & 0.595  &  7.978e+02 &  7.000e+00 &  9.350e-03 &  1.954e-01 &  8.827e+00 &  1.380e-02 &  6.123e-03 \\ 
$5/2^{+}_{4}$ & 4.053 & 0.645  &            &  1.860e+01 &            &  4.205e-02 &  7.821e-01 &  8.911e-03 &  2.937e-03 \\ 
$5/2^{+}_{5}$ & 4.227 & 0.819  &            &  1.213e+02 &            &  9.600e-04 &  1.164e-01 &  8.966e-03 &  2.500e-03 \\ 
$3/2^{+}_{5}$ & 4.453 & 1.045  &            &  6.283e+02 &            &  5.934e-02 &  3.728e+01 &  1.327e-01 &  2.315e-02 \\ 
$11/2^{+}_{1}$ & 4.540 & 1.132  &            &  8.400e+02 &            &  3.496e-01 &  2.937e+02 &  5.662e-03 &  3.775e-03 \\ 
$7/2^{+}_{4}$ & 4.657 & 1.249  &  6.840e+04 &  1.600e+03 &  1.990e-03 &  1.536e-02 &  1.607e+02 &  5.464e-02 &  2.348e-02 \\ 
$7/2^{+}_{5}$ & 4.940 & 1.532  &  5.500e+04 &  5.700e+03 &  5.820e-03 &  3.131e-01 &  2.104e+03 &  3.028e-02 &  1.345e-02 \\ 
\hline\hline
$7/2^{+}_{3}$ & 4.003 & 0.169  &           &  4.000e-06 &           &  5.820e-03 &  2.328e-08 &  1.380e-02 &  4.844e-11 \\ 
$1/2^{+}_{2}$ & 4.048 & 0.214  & 1.690e-02 &  9.360e-05 & 4.028e-01 &  1.700e-03 &  6.808e-03 &  4.066e-01 &  2.232e-03 \\ 
$5/2^{+}_{4}$ & 4.053 & 0.219  &           &  1.641e-04 &            &  2.388e-02 &  3.919e-06 &  8.911e-03 &  4.414e-08 \\ 
$5/2^{+}_{5}$ & 4.227 & 0.393  &           &  1.410e-01 &            &  9.805e-02 &  1.383e-02 &  8.966e-03 &  8.903e-04 \\ 
$3/2^{+}_{5}$ & 4.453 & 0.619  & 1.000e+03 &  1.000e+01 &  9.720e-03 &  3.590e-02 &  1.008e+01 &  1.327e-01 &  1.878e-02 \\ 
$7/2^{+}_{4}$ & 4.657 & 0.823  &           &  1.277e+02 &            &  4.277e-02 &  5.462e+00 &  5.464e-02 &  2.394e-03 \\ 
$7/2^{+}_{5}$ & 4.940 & 1.106  &           &  9.390e+02 &            &  4.700e-04 &  4.413e-01 &  3.028e-02 &  8.465e-06 \\ 
\end{tabular}
\end{ruledtabular}
\end{table*}

$^{24}$Al is an odd--odd nucleus, $T=1$. It has a rather low--lying first excited state, spin $1^+$, at 425.8 keV. Spin of the ground state is $4^+$. There is also close lying $2^+$ state, at 510 keV. USD shell model gives estimates for energies of 447 keV and 588 keV respectfully. Proton separation energy of $^{25}$Si is 3.408 MeV. Spins of expermental excitation energies in $^{25}$Si are not known, therefore Shell model energies are used.

\begin{figure*}[htbp]
\includegraphics[width=8cm]{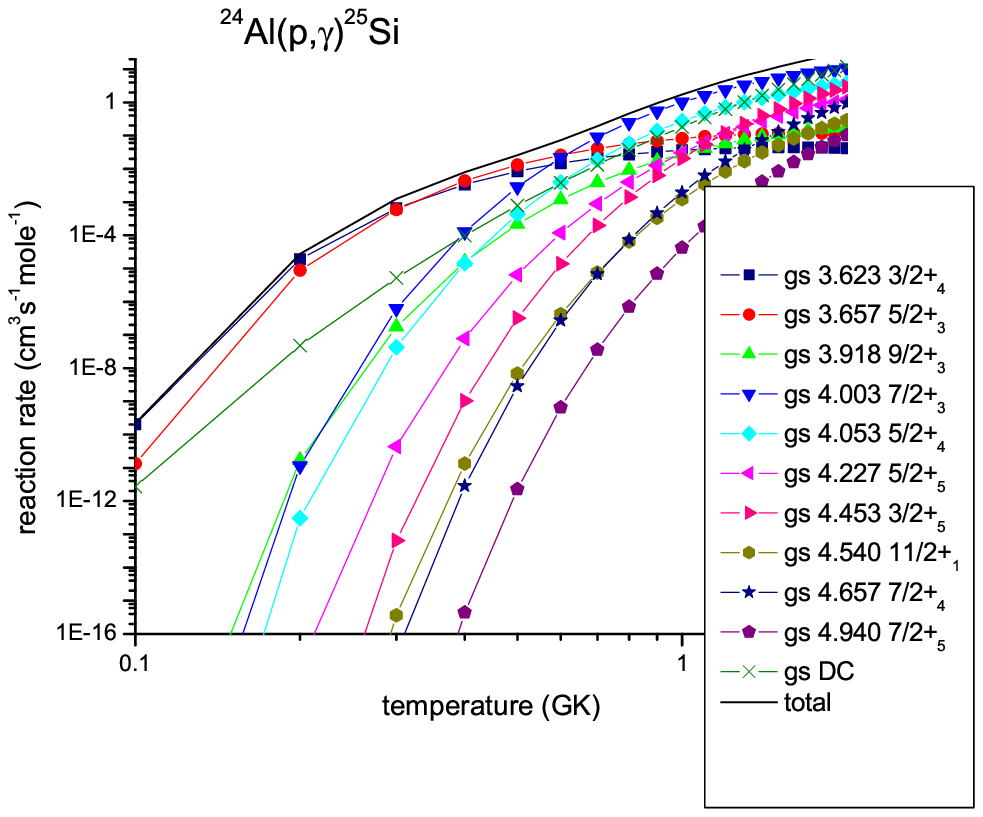}
\includegraphics[width=8cm]{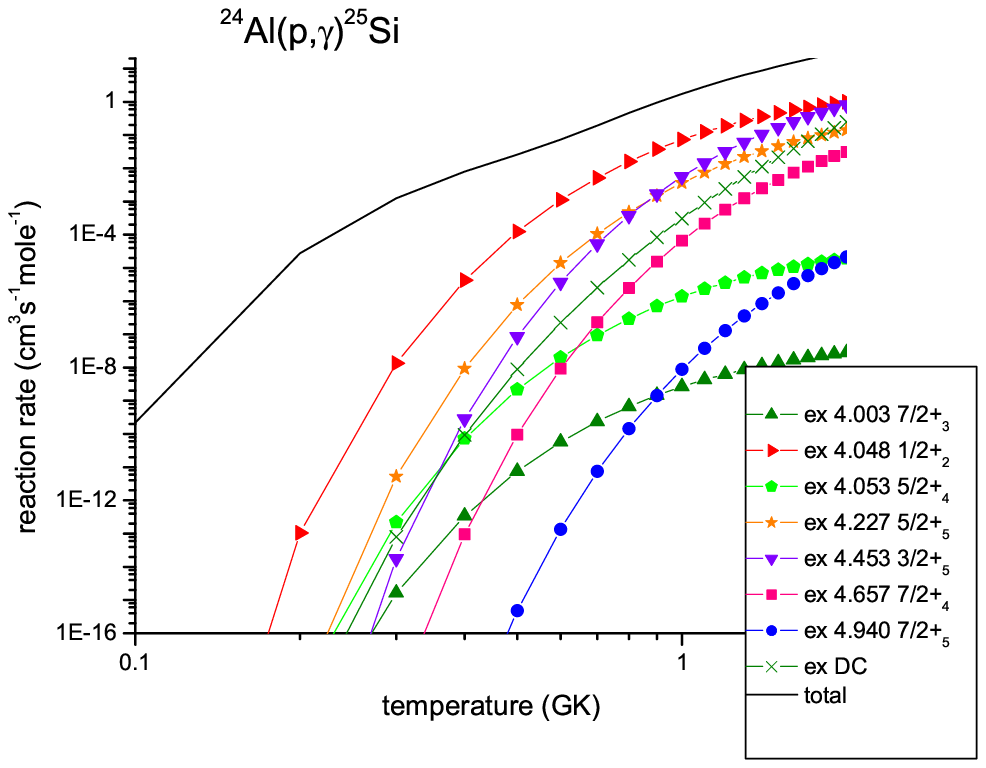}
\caption{\label{24res} The contributions of various individual resonances and through direct capture (DC) to the $^{24}$Al(p,$\gamma$)$^{25}$Si reaction rate as functions of temperature. In the legend, resonances are labeled with their excitation energy in $^{25}$Si. The right panel shows contributions from capture on the ground state in $^{24}$Al, $J^\pi=4^+$, the left panel contributions from capture on the first excited state, $J^\pi=1^+$, in each case weighted with relative population of the respective target state. Both panels show the same total $^{24}$Al(p,$\gamma$)$^{25}$Si reaction rate for comparison.}
\end{figure*}

\begin{figure}[htbp]
\includegraphics[width=8cm]{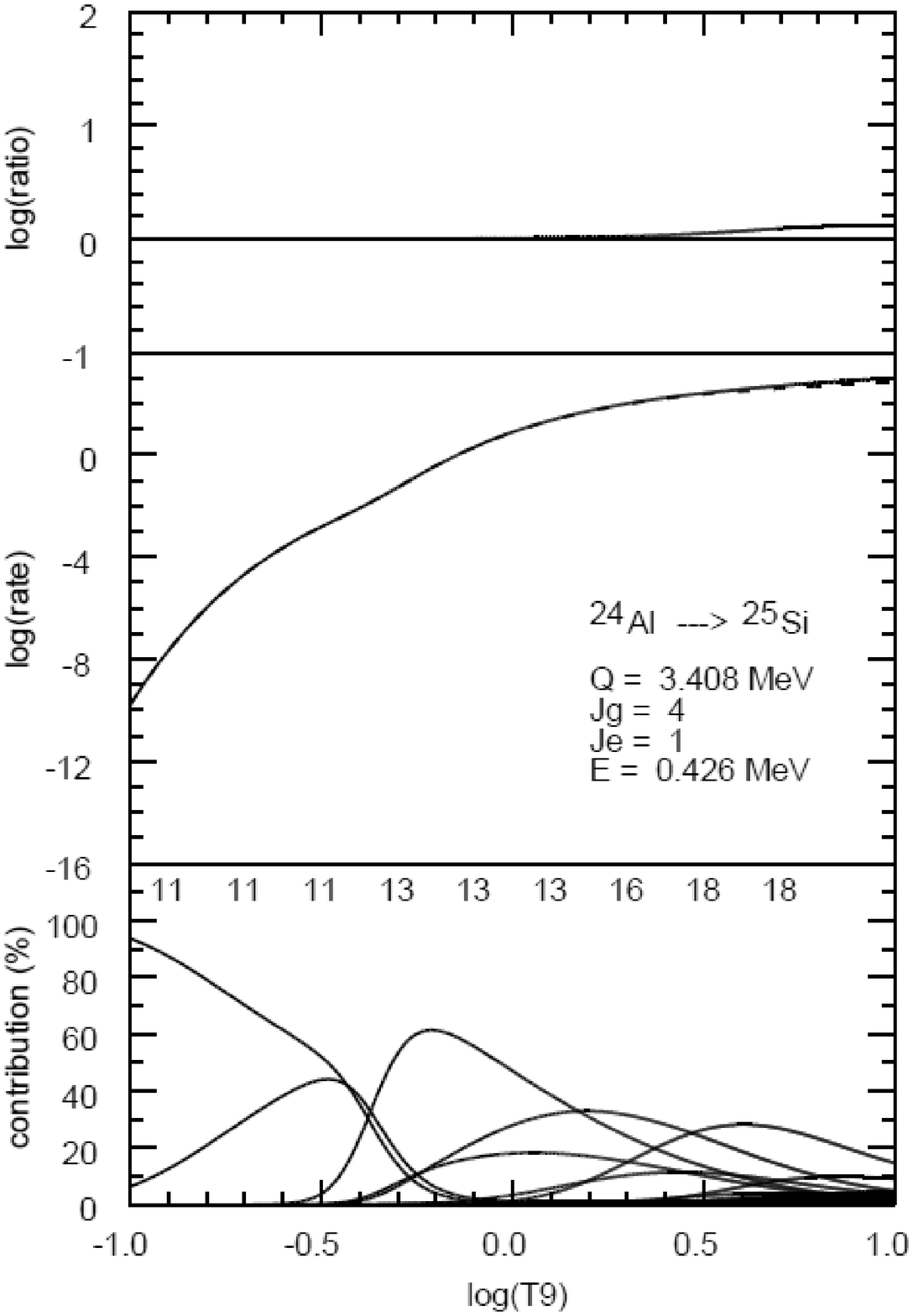}
\caption{\label{Fig4} Results for $^{24}$Al$(4^{+},1^{+}) \rightarrow ^{25}$Si. The middle panel shows the rp-reaction rate for the $^{24}$Al $4^{+}$ ground state (dashed line) and the sum of the contributions from the ground and a $1^{+}$ excited state at 0.426  MeV (full line). See caption to Fig. \ref{Fig1} for other details.}
\end{figure}

Total $^{24}$Al(p,$\gamma$)$^{25}$Si reaction rate dominated by several resonances for proton capture on a ground state, as shown in Fig.~\ref{24res} and Fig. \ref{Fig4}.

\subsection{$^{34}$Cl(p,$\gamma$)$^{35}$Ar}

\begin{table*}
\caption{\label{34ClDC} Spectroscopic factors $C^2S$ and astrophysical S-factors $S(E_0)$ for direct capture into bound states in $^{35}$Ar for temperature range of $0.1-2.0$ GK. Listed are results for capture on the $^{34}$Cl ground state, $J^\pi=0^+$, as well as on the first excited state in $^{34}$Cl, $J^\pi=3^+$ (denoted with an asterix). $J^\pi$ are spin and parity of the $^{35}$Ar final state, $n$ is the node number, $l_0$ the single particle orbital momentum, and $j_0$ the total single particle angular momentum. $E_x$ is excitation energy.}
\begin{ruledtabular}
\begin{tabular}{dccdcdc}
 E_x {\rm ( MeV)} & $J^\pi$ & $(nl_0)_{j0}$ & C^2S & $S(E_0) {\rm ( MeV barn)}$ &  C^2S^* & $S(E_0)^* {\rm ( MeV barn)}$ \\
0     & $3/2^+$ & 1d$_{3/2}$   & 0.954  & 3.13e-1 -- 3.28e+0     & 1.050     & 5.23e-2 -- 1.32e+0 \\
      &         & 1d$_{5/2}$   &        &                        & 0.001     & 9.26e-5 -- 8.73e-3 \\
1.243 & $1/2^+$ & 2s$_{1/2}$   & 0.143  & 2.34e-1 -- 5.95e+0     &           &                    \\
      &         & 1d$_{5/2}$   &        &                        & 0.006     & 2.92e-4 -- 2.60e-2 \\ 
1.780 & $5/2^+$ & 2s$_{1/2}$   &        &                        & 0.012     & 7.85e-3 -- 4.82e-1 \\
      &         & 1d$_{3/2}$   &        &                        & 0.328     & 1.69e-2 -- 3.67e-1 \\
      &         & 1d$_{5/2}$   & 0.006  & 5.08e-3 -- 1.72e-1     & 0.012     & 1.50e-3 -- 1.30e-1 \\
2.671 & $3/2^+$ & 1d$_{3/2}$   & 0.011  & 1.97e-3 -- 1.65e-2     &           &                    \\
2.778 & $7/2^+$ & 2s$_{1/2}$   &        &                        & 0.001     & 3.68e-4 -- 1.92e-2 \\
      &         & 1d$_{3/2}$   &        &                        & 0.662     & 3.50e-2 -- 6.78e-1 \\
      &         & 1d$_{5/2}$   &        &                        & 0.003     & 3.54e-4 -- 2.87e-2 \\ 
3.145 & $5/2^+$ & 2s$_{1/2}$   &        &                        & 0.003     & 1.34e-3 -- 6.44e-2 \\
      &         & 1d$_{3/2}$   &        &                        & 0.017     & 6.02e-4 -- 1.11e-2 \\
      &         & 1d$_{5/2}$   & 0.035  & 1.73e-2 -- 5.32e-1     & 0.006     & 4.82e-4 -- 3.79e-2 \\
3.962 & $3/2^+$ & 1d$_{3/2}$   & 0.027  & 3.11e-3 -- 2.21e-2     &           &                    \\
      &         & 1d$_{5/2}$   &        &                        & 0.002     & 8.25e-5 -- 5.98e-3 \\
4.087 & $1/2^+$ & 2s$_{1/2}$   & 0.023  & 1.89e-2 -- 2.63e-1     &           &                    \\
4.196 & $9/2^+$ & 1d$_{5/2}$   &        &                        & 0.015     & 1.24e-3 -- 8.75e-2 \\
4.405 & $7/2^+$ & 2s$_{1/2}$   &        &                        & 0.068     & 3.12e-2 -- 9.78e-1 \\
      &         & 1d$_{3/2}$   &        &                        & 0.346     & 1.03e-2 -- 1.53e-1 \\
      &         & 1d$_{5/2}$   &        &                        & 0.008     & 4.73e-4 -- 3.25e-2 \\
4.595 & $5/2^+$ & 2s$_{1/2}$   &        &                        & 0.005     & 1.60e-3 -- 4.56e-2 \\
      &         & 1d$_{3/2}$   &        &                        & 0.281     & 5.75e-3 -- 8.23e-2 \\
      &         & 1d$_{5/2}$   & 0.001  & 1.96e-4 -- 5.17e-3     & 0.005     & 1.96e-4 -- 1.31e-2 \\ 
4.778 & $3/2^+$ & 1d$_{3/2}$   & 0.003  & 2.43e-4 -- 1.47e-3     & 0.001     & 1.16e-5 -- 1.59e-4 \\
5.088 & $1/2^+$ & 2s$_{1/2}$   & 0.001  & 3.83e-4 -- 3.26e-3     &           &                    \\
      &         & 1d$_{5/2}$   &        &                        & 0.001     & 1.18e-5 -- 7.31e-4 \\
5.120 & $7/2^+$ & 1d$_{3/2}$   &        &                        & 0.005     & 1.09e-4 -- 1.39e-3 \\
5.535 & $5/2^+$ & 2s$_{1/2}$   &        &                        & 0.005     & 1.28e-3 -- 2.07e-2 \\
      &         & 1d$_{3/2}$   &        &                        & 0.007     & 8.82e-5 -- 1.02e-3 \\
      &         & 1d$_{5/2}$   & 0.010  & 1.24e-3 -- 2.85e-2     & 0.006     & 1.19e-4 -- 6.79e-3 \\
5.572 & $3/2^+$ & 1d$_{3/2}$   & 0.364  & 1.79e-2 -- 9.90e-2     &           &                    \\
5.676 & $3/2^+$ & 1d$_{3/2}$   & 0.013  & 5.74e-4 -- 3.14e-3     &           &                    \\
5.878 & $5/2^+$ & 2s$_{1/2}$   &        &                        & 0.062     & 1.50e-2 -- 1.87e-1 \\
      &         & 1d$_{3/2}$   &        &                        & 0.032     & 2.82e-4 -- 3.07e-3 \\
      &         & 1d$_{5/2}$   & 0.002  & 1.75e-4 -- 4.10e-3     & 0.011     & 1.66e-4 -- 9.04e-3 \\ \hline
\mbox{total} &    &            &        & 6.13e-1 -- 1.04e+1     &           & 1.93e-1 -- 4.80e+0 \\
\end{tabular}
\end{ruledtabular}
\end{table*}

\begin{table*}[htbp] 
\caption{\label{34Cl}Properties of resonant states for $^{34}$Cl$(p,\gamma)$ reaction. Listed are spin and parity $J^\pi$, excitation energy $E_x$, center of mass resonance energy $E_r$, proton single particle widths $\Gamma_{\rm sp}$ for angular momenta $l$, spectroscopic factors $C^2S$, proton-decay width $\Gamma_p$, $\gamma$-decay width $\Gamma_\gamma$ and the
resonance strength $\omega\gamma$. The upper part is for ground state capture, $J^\pi=0^+$, the lower part for capture on the first excited state in $^{34}$Cl, $J^\pi=3^+$. The isospin of $^{33}$Ar states listed is $1/2$, except for the $1/2^+$ state at 7.514$ MeV$ with $T=3/2$}
\begin{ruledtabular}
\begin{tabular}{cccccccccc}
$J^\pi$  & $E_x {\rm ( MeV)}$ & $ E_r {\rm ( MeV)}$ & \multicolumn{2}{c}{$\Gamma_{\rm sp}$}  & \multicolumn{2}{c}{$C^2S$} & $\Gamma_{\rm p} {\rm (eV)}$ & $\Gamma_\gamma {\rm (eV)}$ & $\omega\gamma {\rm (eV)}$ \\
 & & & l=0 & l=2 & l=0 & l=2 \\  
$5/2^{+}_{6}$ & 6.394 & 0.497  &            &  4.830e-02 &           &  2.910e-03 &  1.406e-04 &  7.392e-01 &  4.216e-04 \\ 
$3/2^{+}_{6}$ & 6.932 & 1.035  &            &  6.260e+01 &           &  1.550e-03 &  9.703e-02 &  9.429e-01 &  1.736e-01 \\ 
$5/2^{+}_{7}$ & 7.049 & 1.152  &            &  2.016e+02 &            &  9.000e-05 &  1.814e-02 &  6.259e-01 &  3.576e-02 \\ 
$1/2^{+}_{4}$ & 7.208 & 1.311  &  1.670e+04 &            &  1.140e-03 &            &  1.904e+01 &  2.042e+00 &  1.764e+00 \\ 
$1/2^{+}_{5}$ & 7.404 & 1.507  &  4.070e+04 &            & 1.059e-01 &             &  4.310e+03 &  7.554e-02 &  7.554e-02 \\ 
$1/2^{+}_{1, {\rm T=3/2}}$ & 7.514 & 1.617  & 5.880e+04 &       & 1.127e-01 &      &  6.629e+03 &  3.518e-02 &  3.518e-02 \\ 
$3/2^{+}_{7}$ & 7.577 & 1.680  &            &  2.600e+03 &            &  2.800e-03 &  7.280e+00 &  1.561e+00 &  1.352e+00 \\  
\hline\hline
$7/2^{+}_{4}$ & 6.221 & 0.178  & 4.100e-07 &  4.100e-09 & 2.741e-02 &  2.550e-02 &  1.134e-08 &  4.591e-02 &  6.482e-09 \\ 
$5/2^{+}_{6}$ & 6.394 & 0.351  & 3.850e-02 &  4.015e-04 & 1.110e-03 &  1.340e-03 &  4.327e-05 &  7.392e-01 &  1.854e-05 \\ 
$9/2^{+}_{2}$ & 6.695 & 0.652  &           &  8.170e-01 &           &  1.144e-02 &  9.364e-03 &  8.257e-02 &  5.997e-03 \\ 
$7/2^{+}_{5}$ & 6.883 & 0.840  & 5.950e+02 &  1.360e+01 & 1.420e-03 &  1.527e-02 &  1.053e+00 &  2.487e-01 &  1.149e-01 \\ 
$3/2^{+}_{6}$ & 6.932 & 0.889  &           &  1.720e+01 &            &  8.200e-04 &  1.410e-02 &  9.429e-01 &  3.605e-03 \\ 
$5/2^{+}_{7}$ & 7.049 & 1.006  & 2.700e+03 &  6.610e+01 &  7.000e-05 &  1.810e-03 &  3.086e-01 &  6.259e-01 &  8.690e-02 \\ 
$1/2^{+}_{4}$ & 7.208 & 1.165  &           &  2.142e+02 &            &  4.480e-03 &  9.596e-01 &  2.042e+00 &  1.270e-02 \\ 
$9/2^{+}_{3}$ & 7.240 & 1.197  &           &  2.043e+02 &            &  2.810e-03 &  5.741e-01 &  1.615e-01 &  9.003e-02 \\ 
$1/2^{+}_{5}$ & 7.404 & 1.361  &           &  7.583e+02 &            &  6.000e-05 &  4.550e-02 &  7.554e-02 &  1.139e-07 \\ 
$3/2^{+}_{7}$ & 7.577 & 1.534  &           &  1.300e+03 &            &  6.130e-03 &  7.969e+00 &  1.561e+00 &  2.114e-01 \\ 
\end{tabular}
\end{ruledtabular}
\end{table*}

The first excited state of $^{34}$Cl, spin $3^+$, is at 146.36 keV. Spin of the ground state is $0^+$. The second excited state, spin $1^+$, is at 461 keV. USD shell model gives energies of 133 keV and 317 keV. Proton separation energy of $^{35}$Ar is 5.897 MeV, $T_z=0$. Spins of experimental excitation energies of $^{35}$Ar are not well known except the first $3/2^+$ state with $T=3/2$, therefore for all other states Shell model energies are used. 

\begin{figure*}[htbp]
\includegraphics[width=8cm]{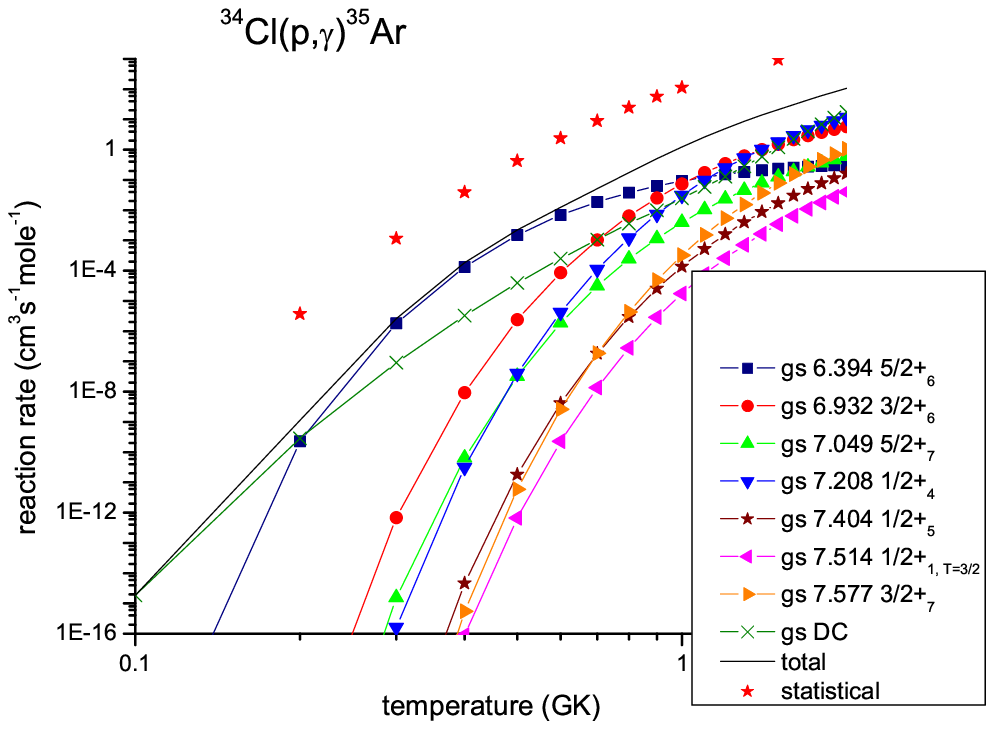}
\includegraphics[width=8cm]{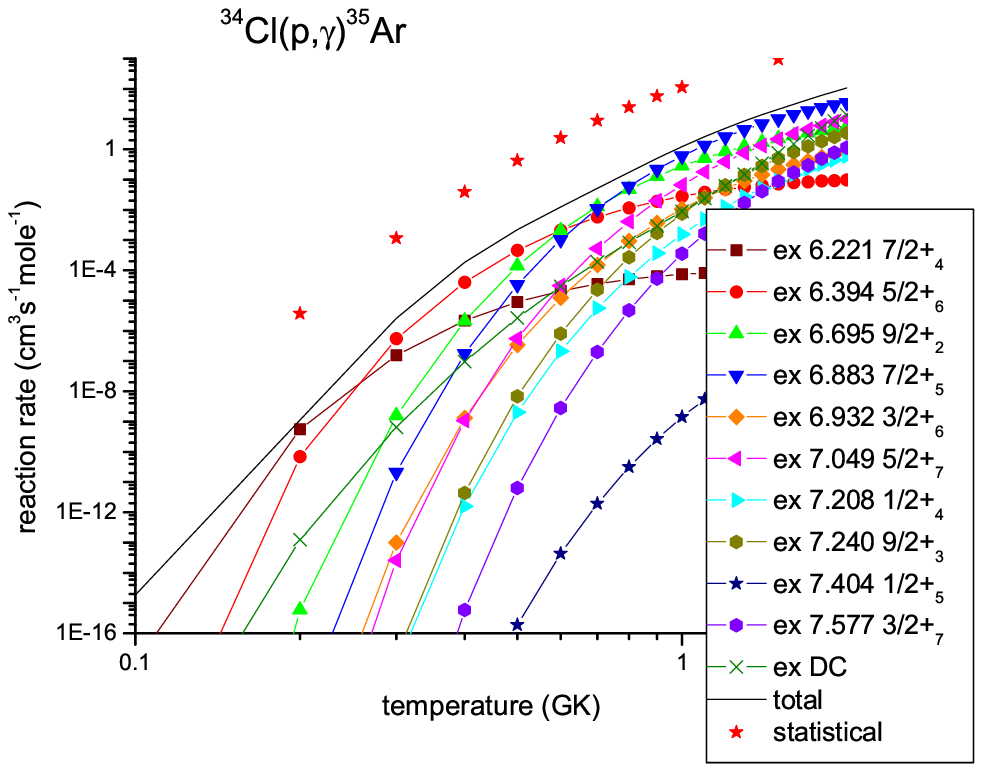}
\caption{\label{34res} The contributions of various individual resonances and through direct capture (DC) to the $^{34}$Cl(p,$\gamma$)$^{35}$Ar reaction rate as functions of temperature. In the legend, resonances are labeled with their excitation energy in $^{35}$Ar. The right panel shows contributions from capture on the ground state in $^{34}$Cl, $J^\pi=0^+$, the left panel contributions from capture on the first excited state, $J^\pi=3^+$, in each case weighted with relative population of the respective target state. Both panels show the same total $^{34}$Cl(p,$\gamma$)$^{35}$Ar reaction rate for comparison. Prediction of statistical model reaction rate, taken from Iliadis et al. \cite{Iliadis01} is also given.}
\end{figure*}

\begin{figure}[htbp]
\includegraphics[width=8cm]{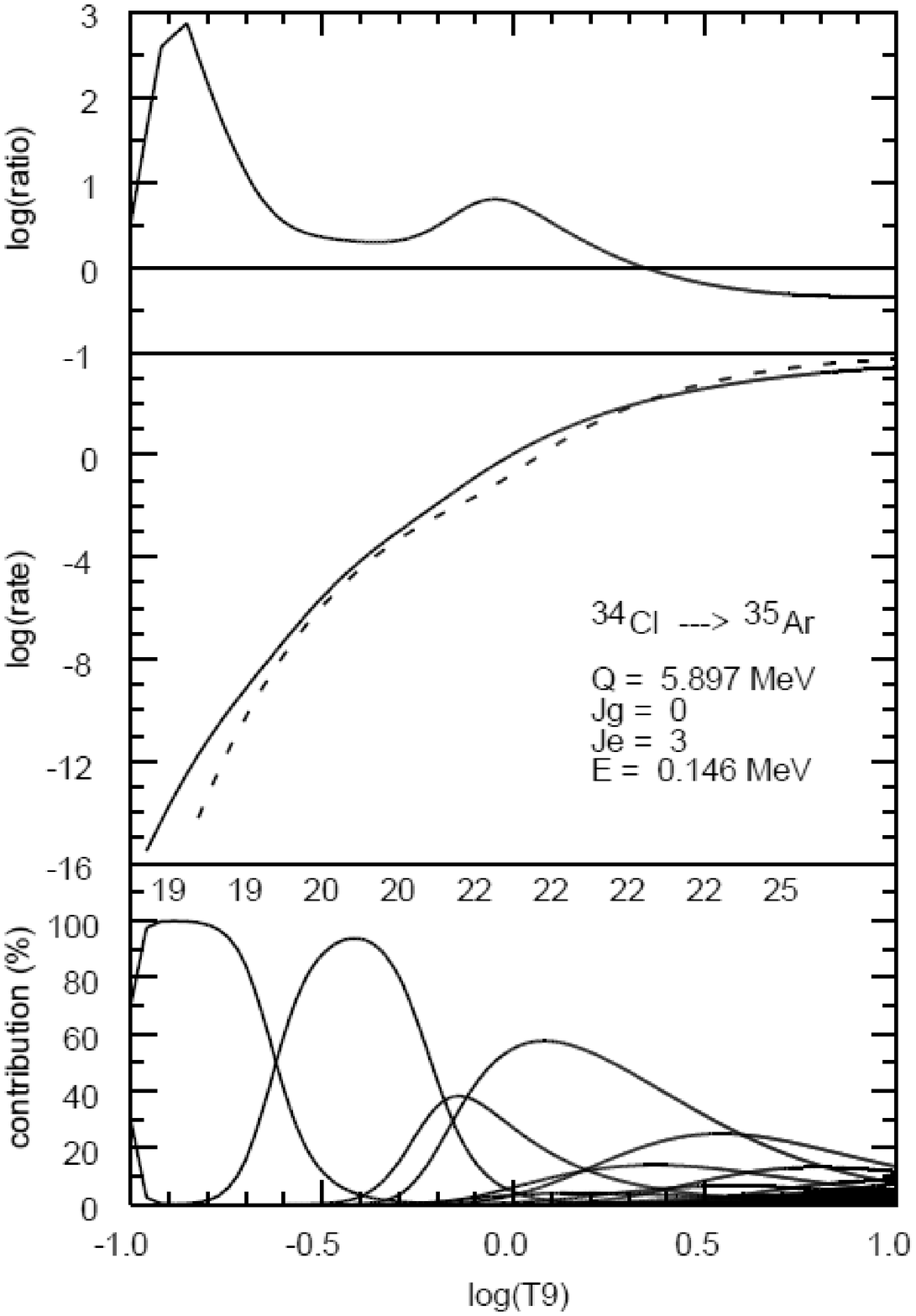}
\caption{\label{Fig2} Results for $^{34}$Cl$(0^{+},3^{+}) \rightarrow ^{35}$Ar. The middle panel shows the rp-reaction rate for the $^{34}$Cl $0^{+}$ ground state (dashed line) and the sum of the contributions from the ground and the $3^{+}$ excited state at 0.146  MeV (full line). See caption to Fig. \ref{Fig1} for other details.}
\end{figure}

As seen from Fig. \ref{34res} and Fig. \ref{Fig2}, total proton capture rate is dominated by resonant capture on a ground state of $^{34}$Cl and a few resonances on the excited state. The first excited state of $^{34}$Cl gives rise to an extremely large SEF of $10^{3}$ at low temperatures peaking at 0.2  GK. This is due to state (19,7/2$^{+}$,4) at 6.222  MeV that is connected by a small $l=0$ spectroscopic factor to the $3^{+}$ state. At 1  GK there is a SEF factor of about 10 from the (22,7/2$^{+}$,5) at 6.884  MeV state that also connects to the $3^{+}$ with a small $l=0$ spectroscopic factor. Unfortunately there is nothing known about these 7/2$^{+}$ states in $^{35}$Ar or it's mirror $^{35}$Cl. Given that the energy of these states have a theoretcal error of the order of 150  keV, we must conclude that there is about three orders of magnitude uncertainty in the $^{34}$Cl(p,$\gamma$)$^{35}$Ar reaction rate.


\subsection{$^{28}$P(p,$\gamma$)$^{29}$S}

\begin{table*}
\caption{\label{28PDC} Spectroscopic factors $C^2S$ and astrophysical S-factors $S(E_0)$ for direct capture into bound states in $^{29}$S for temperature range of $0.1-2.0$ GK. Listed are results for capture on the $^{28}$P ground state, $J^\pi=3^+$, as well as on the first excited state in $^{28}$P, $J^\pi=2^+$ (denoted with an asterix). $J^\pi$ are spin and parity of the $^{29}$S final state, $n$ is the node number, $l_0$ the single particle orbital momentum, and $j_0$ the total single particle angular momentum. $E_x$ is excitation energy.}
\begin{ruledtabular}
\begin{tabular}{dccdcdc}
 E_x {\rm ( MeV)} & $J^\pi$ & $(nl_0)_{j0}$ & C^2S & $S(E_0) {\rm ( MeV barn)}$ &  C^2S^* & $S(E_0)^* {\rm ( MeV barn)}$ \\
0     & $5/2^+$ & 2s$_{1/2}$   & 0.414  & 4.72e-2 -- 6.41e-2     & 0.341     & 5.57e-2 -- 7.80e-2 \\
      &         & 1d$_{3/2}$   & 0.086  & 7.60e-4 -- 9.95e-4     & 0.136     & 1.74e-3 -- 2.31e-3 \\
      &         & 1d$_{5/2}$   & 0.064  & 1.04e-3 -- 1.80e-3     & 0.025     & 5.93e-4 -- 1.05e-3 \\
1.215 & $1/2^+$ & 1d$_{3/2}$   &        &                        & 0.051     & 1.60e-4 -- 2.12e-4 \\
      &         & 1d$_{5/2}$   & 0.057  & 2.16e-4 -- 3.67e-4     & 0.005     & 2.72e-5 -- 4.76e-5 \\ 
1.799 & $7/2^+$ & 2s$_{1/2}$   & 0.013  & 1.49e-3 -- 1.70e-3     &           &                    \\
      &         & 1d$_{3/2}$   & 0.449  & 3.20e-3 -- 4.19e-3     & 0.209     & 2.18e-3 -- 2.89e-3 \\
      &         & 1d$_{5/2}$   & 0.055  & 6.56e-4 -- 1.11e-3     & 0.042     & 7.39e-4 -- 1.29e-3 \\
1.959 & $3/2^+$ & 2s$_{1/2}$   &        &                        & 0.013     & 1.00e-3 -- 1.14e-3 \\
      &         & 1d$_{3/2}$   &        &                        & 0.071     & 3.51e-4 -- 4.67e-4 \\
      &         & 1d$_{5/2}$   & 0.031  & 1.76e-4 -- 2.99e-4     & 0.026     & 2.18e-4 -- 3.79e-4 \\
2.735 & $3/2^+$ & 2s$_{1/2}$   &        &                        & 0.028     & 1.87e-3 -- 2.52e-3 \\
      &         & 1d$_{3/2}$   & 0.002  & 3.60e-6 -- 4.98e-6     & 0.135     & 4.56e-4 -- 6.31e-4 \\
      &         & 1d$_{5/2}$   & 0.022  & 7.71e-5 -- 1.36e-4     & 0.003     & 1.61e-5 -- 2.86e-5 \\ 
2.801 & $5/2^+$ & 2s$_{1/2}$   & 0.065  & 4.38e-3 -- 6.51e-3     & 0.010     & 9.94e-4 -- 1.38e-3 \\
      &         & 1d$_{3/2}$   & 0.058  & 1.86e-4 -- 2.60e-4     & 0.262     & 1.27e-3 -- 1.78e-3 \\
      &         & 1d$_{5/2}$   & 0.001  & 6.52e-6 -- 1.16e-5     & 0.001     & 1.88e-5 -- 2.82e-5 \\
3.017 & $5/2^+$ & 2s$_{1/2}$   & 0.075  & 4.71e-3 -- 8.33e-3     & 0.025     & 2.29e-3 -- 3.66e-3 \\
      &         & 1d$_{3/2}$   & 0.036  & 9.85e-5 -- 1.45e-4     & 0.092     & 3.82e-4 -- 5.53e-4 \\
      &         & 1d$_{5/2}$   &        &                        & 0.001     & 4.86e-6 -- 8.98e-6 \\ \hline
\mbox{total} &    &            &        & 6.76e-2 -- 8.45e-2     &           & 7.23e-2 -- 9.62e-2 \\
\end{tabular}
\end{ruledtabular}
\end{table*}

\begin{table*}[htbp] 
\caption{\label{28P}Properties of resonant states for $^{28}$P$(p,\gamma)$ reaction. Listed are spin and parity $J^\pi$, excitation energy $E_x$, center of mass resonance energy $E_r$, proton single particle widths $\Gamma_{\rm sp}$ for angular momenta $l$, spectroscopic factors $C^2S$, proton-decay width $\Gamma_p$, $\gamma$-decay width $\Gamma_\gamma$ and the
resonance strength $\omega\gamma$. The upper part is for ground state capture, $J^\pi=3^+$, the lower part for capture on the first excited state in $^{28}$P, $J^\pi=2^+$.}
\begin{ruledtabular}
\begin{tabular}{cccccccccc}
$J^\pi$  & $E_x {\rm ( MeV)}$ & $ E_r {\rm ( MeV)}$ & \multicolumn{2}{c}{$\Gamma_{\rm sp}$}  & \multicolumn{2}{c}{$C^2S$} & $\Gamma_{\rm p} {\rm (eV)}$ & $\Gamma_\gamma {\rm (eV)}$ & $\omega\gamma {\rm (eV)}$ \\
 & & & l=0 & l=2 & l=0 & l=2 \\ 
$1/2^{+}_{2}$ & 3.330 & 0.040  &           &  3.200e-24 &           &  5.800e-04 &  1.856e-27 &  4.580e-02 &  2.651e-28 \\ 
$9/2^{+}_{1}$ & 3.381 & 0.091  &           &  1.500e-13 &           &  6.142e-01 &  9.213e-14 &  3.463e-02 &  6.581e-14 \\ 
$3/2^{+}_{3}$ & 3.577 & 0.287  &           &  2.600e-10 &           &  1.193e-02 &  3.102e-12 &  7.339e-02 &  8.862e-13 \\  
$5/2^{+}_{4}$ & 3.817 & 0.527  & 4.540e+01 &  4.231e-01 & 2.080e-03 &  3.944e-01 &  2.613e-01 &  5.081e-02 &  5.075e-03 \\ 
$7/2^{+}_{2}$ & 3.875 & 0.585  & 1.1900e+02 & 1.200e+00 & 3.000e-05 &  1.239e-02 &  1.844e-02 &  2.823e-03 &  1.324e-03 \\ 
$3/2^{+}_{4}$ & 3.977 & 0.687  &           &  5.500e+00 &           &  7.914e-02 &  4.353e-01 &  1.671e-01 &  1.511e-02 \\ 
$5/2^{+}_{5}$ & 4.128 & 0.838  & 2.200e+03 &  3.140e+01 & 1.247e-01 &  3.593e-02 &  2.754e+02 &  4.067e-02 &  1.462e-02 \\ 
$7/2^{+}_{3}$ & 4.446 & 1.156  & 2.010e+04 &  3.888e+02 & 4.840e-03 &  1.507e-02 &  1.031e+02 &  1.850e-02 &  8.696e-03 \\ 
$3/2^{+}_{5}$ & 4.459 & 1.169  &           &  4.242e+02 &           &  1.519e-01 &  6.444e+01 &  3.025e-02 &  1.140e-03 \\ 
$5/2^{+}_{6}$ & 4.501 & 1.211  & 2.390e+04 &  5.672e+02 & 6.250e-03 &  8.530e-03 &  1.542e+02 &  1.988e-01 &  8.041e-02 \\ 
$7/2^{+}_{4}$ & 4.623 & 1.333  & 4.070e+04 &  1.100e+03 & 1.103e-02 &  2.507e-01 &  7.247e+02 &  9.207e-02 &  4.092e-02 \\
$1/2^{+}_{4}$ & 4.650 & 1.360  &           &  1.600e+03 &           &  3.072e-02 &  4.915e+01 &  5.014e-02 &  4.505e-03 \\
\hline\hline
$3/2^{+}_{3}$ & 3.577 & 0.181  & 3.230e-05 &  2.100e-07 & 6.290e-02 &  2.021e-01 &  2.074e-06 &  7.339e-02 &  8.296e-07 \\ 
$5/2^{+}_{4}$ & 3.817 & 0.421  & 4.300e+00 &  3.730e-02 & 1.877e-01 &  5.237e-02 &  8.089e-01 &  5.081e-02 &  1.467e-02 \\ 
$7/2^{+}_{2}$ & 3.875 & 0.479  & 1.680e+01 &  1.554e-01 &           &  7.720e-03 &  1.200e-03 &  2.823e-03 &  1.206e-04 \\ 
$3/2^{+}_{4}$ & 3.977 & 0.581  & 1.048e+02 &  1.300e+00 & 6.770e-03 &  4.871e-02 &  7.728e-01 &  1.671e-01 &  3.756e-02 \\ 
$5/2^{+}_{5}$ & 4.128 & 0.732  & 8.477e+02 &  1.030e+01 & 6.145e-02 &  7.104e-02 &  5.282e+01 &  4.067e-02 &  3.926e-03 \\ 
$1/2^{+}_{3}$ & 4.229 & 0.833  &           &  3.020e+01 &           &  2.602e-01 &  7.858e+00 &  2.903e-02 &  5.785e-03 \\
$7/2^{+}_{3}$ & 4.446 & 1.050  &           &  2.063e+02 &           &  1.077e-01 &  2.221e+01 &  1.850e-02 &  2.622e-03 \\ 
$3/2^{+}_{5}$ & 4.459 & 1.063  & 1.260e+04 &  2.043e+02 & 3.291e-02 &  4.656e-02 &  4.242e+02 &  3.025e-02 &  1.050e-02 \\ 
$5/2^{+}_{6}$ & 4.501 & 1.105  & 1.140e+04 &  2.504e+02 & 6.100e-04 &  8.070e-03 &  8.975e+00 &  1.988e-01 &  6.552e-03 \\ 
$7/2^{+}_{4}$ & 4.623 & 1.227  &           &  5.950e+02 &           &  3.477e-01 &  2.069e+02 &  9.207e-02 &  1.636e-02 \\
$1/2^{+}_{4}$ & 4.650 & 1.254  &           &  6.866e+02 &           &  4.216e-02 &  2.895e+01 &  5.010e-02 &  3.715e-03 \\ 
\end{tabular}
\end{ruledtabular}
\end{table*}

$^{28}$P is also an odd-odd nucleus, $T=1$ . The first excited state of $^{28}$P, spin $2^+$, is at 105.64 keV. Spin of the ground state is $3^+$. The second excited state is at 877  keV and spin $0^+$ is assigned from USD shell model prediction. USD shell model gives energies of 142  keV and 840  keV. Proton separation energy of $^{29}$S is 3.290  MeV.

As seen from Fig. \ref{34res} and Fig. \ref{Fig3}, $^{28}$P(p,$\gamma$)$^{29}$S reaction rate is dominated by fourth $5/2^+$ state resonance at 3.817  MeV (USD shell model energy prediction). Proton capture on a ground state and first excited state of $^{28}$P is equally important. Fig. \ref{Fig3} shows the SEF of about 20 around 0.8  GK due to first excited state. This comes from the final state (11,5/2$^{+}$,4) at 3.817  MeV in $^{29}$S. It is connected to the $2^{+}$ excited state with a relatively strong $l=0$ spectroscopic factor. There are no known excited states in $^{29}$S, but in the mirror nucleus $^{29}$Al, the fourth 5/2$^{+}$ state at 3.641  MeV is in reasonable agreement with the theoretical energy. The analogue energy of this state in $^{29}$S should be within 100  keV of 3.64  MeV, but this is enough to give an order of magnitude change in the rp-process reaction rate around 0.8  GK.

Reaction rate at relevant temperature range $T = 0.1-2$ GK agrees well with statistical model \cite{Iliadis01}, however for higher stellar temperatures up to 10  GK direct capture starts to dominate reaction rate and total reaction rate exceeds statistical model prediction.

\begin{figure*}[htbp]
\includegraphics[width=8cm]{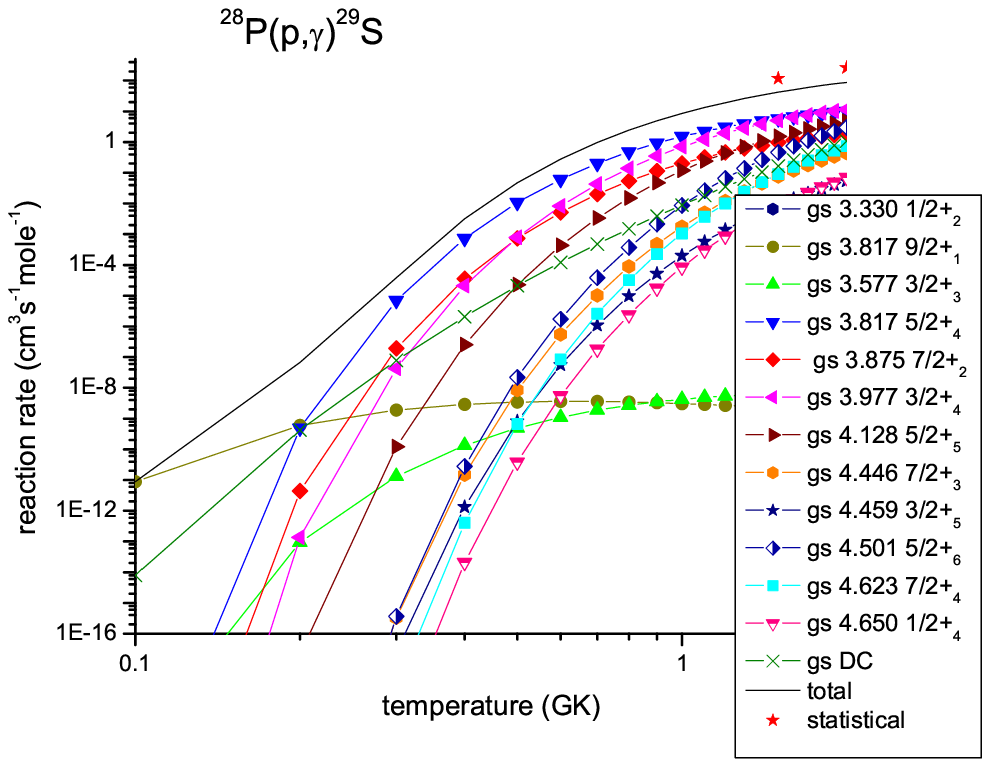}
\includegraphics[width=8cm]{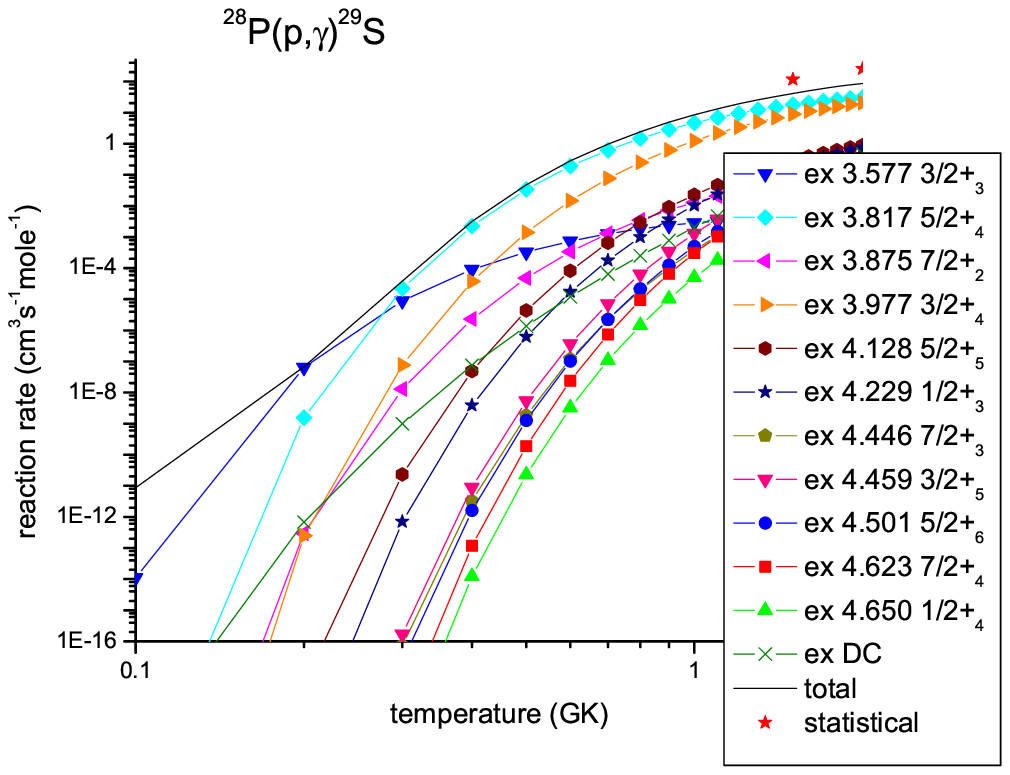}
\caption{\label{28res} The contributions of various individual resonances and through direct capture (DC) to the $^{28}$P(p,$\gamma$)$^{29}$S reaction rate as functions of temperature. In the legend, resonances are labeled with their excitation energy in $^{29}$S. The right panel shows contributions from capture on the ground state in $^{28}$P, $J^\pi=3^+$, the left panel contributions from capture on the first excited state, $J^\pi=2^+$, in each case weighted with relative population of the respective target state. Both panels show the same total $^{28}$P(p,$\gamma$)$^{29}$S reaction rate for comparison. Prediction of statistical model reaction rate, taken from Iliadis et al. \cite{Iliadis01} is also given.}
\end{figure*}

\begin{figure}[htbp]
\includegraphics[width=8cm]{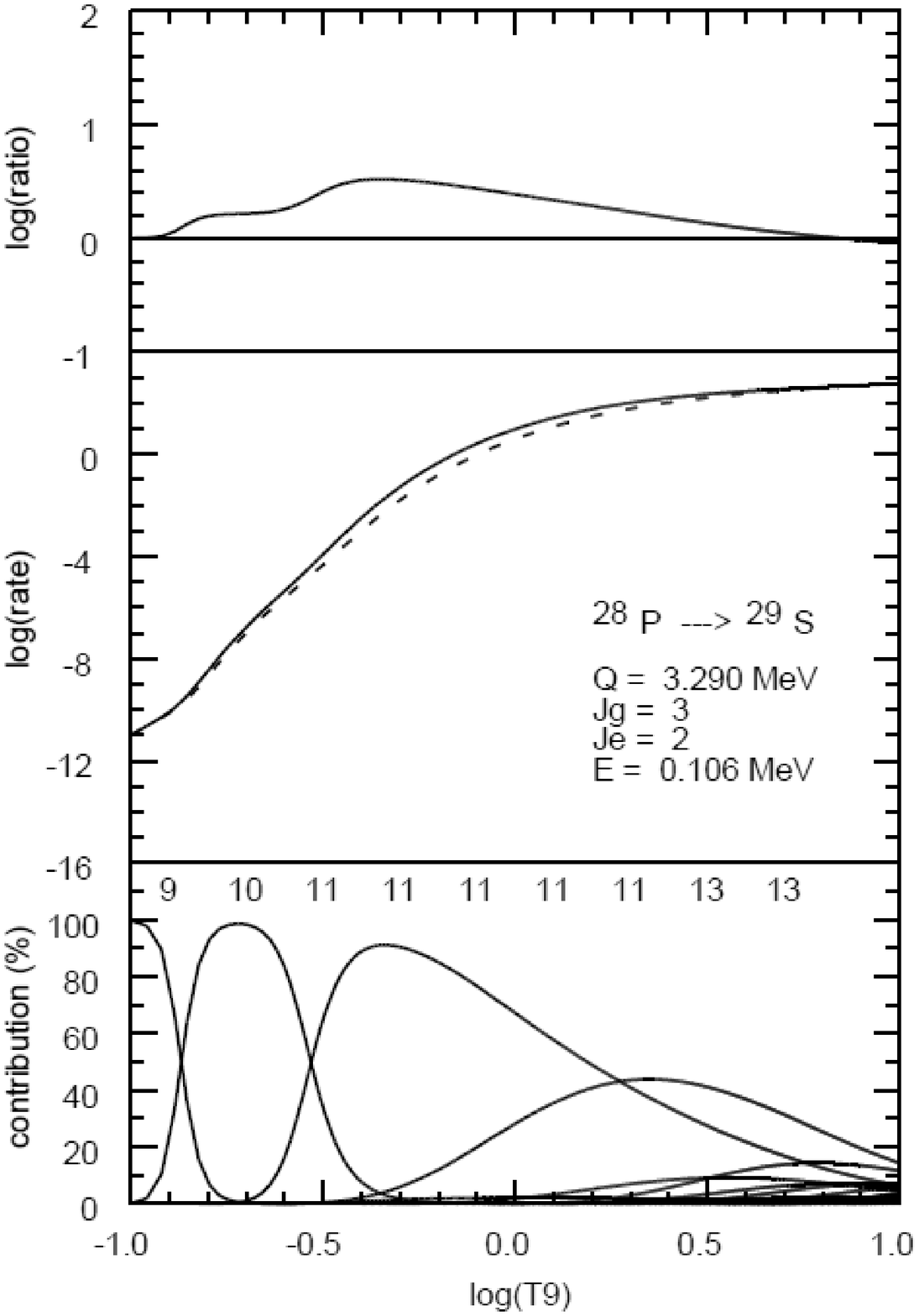}
\caption{\label{Fig3} Results for $^{28}$P$(3^{+},2^{+}) \rightarrow ^{29}$S. The middle panel shows the rp-reaction rate for the $^{28}$P $3^{+}$ ground state (dashed line) and the sum of the contributions from the ground and the $2^{+}$ excited state at 0.106  MeV (full line). See caption to Fig. \ref{Fig1} for other details.}
\end{figure}

\section{Conclusion}

Although very few $sd$ shell nuclei have low--lying excited states, contributing significantly to total rates, proton capture on a first excited state cannot be neglected. In addition to the previously studied case of  $^{32}$Cl$(p,\gamma)^{33}$Ar, we find at least two other cases $^{28}$P$(p,\gamma)^{29}$S and $^{34}$Cl$(p,\gamma)^{35}$Ar for which the low-lying excited state in initial nucleus dominates the cross section. Unfortunately, the excited states are not known for these cases. Initial states above about 300  keV are not important unless there is a particularly strong spectroscopic factor that connects to low-lying final states in the daughter.

Direct capture rates depend on astrophysical S--factor, which changes slightly with temperature at temperatures relevant to rp--process for X--ray bursts, $T=0.1-2$ GK, and may have resonances for higher temperatures. Therefore knowing at what energies resonances exist is important for quality of such reaction rate predictions and so for isotopic abundances and energy release.

An important task for the near future is to quantify the errors associated with these and other cases where the energies and spectroscopic factors are not known experimentally and must be taken from theory. Recent efforts have been me to establish improved Hamiltonians for the sd-shell. Calculation of the corss section with several realistic Hamiltonians will help to establish theoretical error bands for important cross sections.

\end{document}